\documentclass[
 amsmath,amssymb,
 aps,
 pra,
 twocolumn,
]{revtex4-2}

\usepackage{graphicx}% Include figure files
\usepackage{dcolumn}% Align table columns on decimal point
\usepackage{bm}% bold math

\usepackage{times}
\usepackage{amsmath}
\usepackage{amssymb}
\usepackage{braket}
\newcounter{lastnote}

\begin{document}
\preprint{APS/123-QED}

\title{Ultrafast Superconducting Qubit Readout with the Quarton Coupler} 

\author{Yufeng Ye$^{1,2}$, Jeremy B. Kline$^{1,2}$, Sean Chen$^{1,2}$, Kevin P. O'Brien$^{1,2}$}

\email[Correspondence email address: ]{kpobrien@mit.edu}%

\affiliation{$^{1}$Department of Electrical Engineering and Computer Science}
\affiliation{$^{2}$Research Laboratory of Electronics, Massachusetts Institute of Technology, Cambridge, MA 02139, USA}

\date{\today} 

\begin{abstract}

  Fast, high-fidelity, and quantum nondemolition (QND) qubit readout is an essential element of quantum information processing. 
  For superconducting qubits, state-of-the-art readout is based on a dispersive cross-Kerr coupling between a qubit and its readout resonator. The resulting readout can be high-fidelity and QND, but readout times are currently limited to the order of 50 ns due to the dispersive cross-Kerr of magnitude 10 MHz.
  Here, we present a new readout scheme that uses the quarton coupler to facilitate a large (greater than 250 MHz) cross-Kerr between a transmon qubit and its readout resonator. 
  Full master equation simulations show a 5 ns readout time with greater than $99$\% readout and QND fidelity. Unlike state-of-the-art dispersive readout, the proposed ``quartonic readout'' scheme relies on a transmon with linearized transitions as the readout resonator. Such operational points are found from a detailed theoretical treatment and parameter study of the coupled system.
  The quartonic readout circuit is also experimentally feasible and preserves the coherence properties of the qubit.
  Our work reveals a new path for order-of-magnitude improvements of superconducting qubit readout by engineering nonlinear light-matter couplings in parameter regimes unreachable by existing designs.  

\end{abstract}

\maketitle

\section{Introduction}
Fast and high-fidelity qubit readout is essential for quantum error correction \cite{GoogleSurfaceCode22, BeyondBreakEven23} and other feedback schemes in quantum computing and communication including teleportation \cite{teleportation, teleportZurich13} and state-initialization \cite{HeraldedInitialization1, HeraldedInitialization2}. Superconducting qubits \cite{blais2021, EngineersReview} are a leading material platform for quantum information processing \cite{supremacy} in part due to their reliably fast, high-fidelity, and quantum-non-demolition (QND) readout \cite{FirstDispersiveRO, walter2017, Sunada2023}.
In the state-of-the-art dispersive readout \cite{FirstDispersiveRO, Sunada2023} for superconducting qubits, a cross-Kerr interaction $2\chi \hat{a}^\dagger \hat{a} \hat{b}^\dagger \hat{b}$ between a qubit ($\hat{b}^\dagger \hat{b}$) and its auxillary readout resonator ($\hat{a}^\dagger \hat{a}$) entangles the qubit's state with the phase of a driven classical coherent state in the resonator, which then decays with rate $\kappa$ into the environment and is usually measured through heterodyne detection \cite{blais2021, EngineersReview}. 
Compared to non-cross-Kerr readout schemes such as high-power Jaynes-Cummings readout \cite{Reed2010} or longitudinal readout \cite{Didier}, dispersive readout has experimentally demonstrated the fastest readout time (40 ns \cite{Sunada2023}) required for high ($>\!99\%$) readout fidelity \cite{walter2017, Sunada2023} and high ($>\!99\%$) QND fidelity \cite{Sunada2022-IntrinsicPF, Sunada2023}. 

High readout fidelity requires high measurement signal-to-noise-ratio (SNR) \cite{blais2021, bultink2018}. The SNR for cross-Kerr based qubit measurement conveniently scales with \cite{blais2021, bultink2018}:
\begin{equation}
\text{SNR}^2 \propto \eta \kappa \bar{n} 
 |\text{sin}(2\theta)| t,
\end{equation}
where the readout time $t$ required to reach a desired SNR can be reduced by increasing either $ \eta \kappa \bar{n}$, the effective rate at which measurement photons are collected, or $|\text{sin}(2\theta)| = \frac{\chi \kappa}{\chi^2 + \kappa^2/4}$, the amount of phase information each photon carries. Over the past decades, there have been significant advances in designing $2\chi = \kappa$ to maximize $|\text{sin}(2\theta)| = 1$ \cite{Sunada2023} and engineering devices including quantum-limited amplifiers that improve quantum efficiency $\eta$ towards the theoretical maximum $\eta = 1$ \cite{JRM_JPA, TWPAScience, FloquetTWPA}. It has also become better understood that non-idealities in dispersive coupling tends to limit \cite{Sank2016-StateMixing, Chaos-DispersiveReadout} average readout resonator photon number $\bar{n}$ to low values $\bar{n} \lessapprox 5$ \cite{khezri2016, walter2017} when the qubit is a state-of-the-art transmon \cite{KochTransmon}. Notably, a simple way to improve readout SNR is by increasing the coupling rate $\kappa$ of the readout resonator to the environment. While larger $\kappa$ is easily achievable by increasing coupling capacitance, designs with $\kappa/2\pi \gg 10 $ MHz are practically difficult for due to a number of reasons, many of which are ultimately caused by the perturbative nature of dispersive coupling, wherein the nonlinear cross-Kerr interaction is a perturbative effect derived from the underlying qubit-resonator linear coupling $g(\hat{a}^\dagger - \hat{a})(\hat{b}^\dagger - \hat{b})$ \cite{Alexandre2004}. The first reason dispersive readout is incompatible with very large $\kappa$ is the limit on $\chi$. Since $\kappa \approx 2\chi$ is needed for optimizing the term $|\text{sin}(2\theta)|$, larger $\chi$ is needed to accompany larger $\kappa$; but dispersive $\chi$ for state-of-the-art transmon qubits \cite{KochTransmon} with low anharmonicity $E_C/2\pi \approx 200-350$ MHz are limited to $\chi/2\pi \lessapprox 10$ MHz due to the perturbative cross-Kerr being limited to a small fraction of the qubit anharmonicity \cite{KochTransmon}. Secondly, the underlying linear coupling in dispersive coupling causes eigenstates of the qubit-resonator system to be combinations of qubit and readout resonator bare states; so a stronger resonator-environment coupling $\kappa$ invariably increases many qubit eigenstate decoherence rates \cite{beaudoin2011} such as Purcell decay \cite{PurcellDecay}. 
There is thus an opportunity to find a non-perturbative source of cross-Kerr interaction \cite{Dassonneville2020} between superconducting qubits and resonators that allows for a much larger $\chi$ and therefore much larger $\kappa$, which can lead to proportionally larger readout SNR. 

Here, we show in simulation that by designing high $2\chi = \kappa/2\pi \approx 250-300 $ MHz, an order of magnitude higher than state-of-the-art dispersive readout, a cross-Kerr based readout scheme can result in about an order of magnitude faster readout time -- just 5 ns to reach 99\% readout fidelity and QND fidelity. To reach the high $2\chi \approx \kappa$ values, we leverage the quarton coupler we previously proposed in \cite{quartonPRL}, which is capable of ultrastrong $2\chi/2\pi \rightarrow 1 $ GHz cross-Kerr coupling between a transmon qubit and resonator. This resonator is not the typical standing-wave waveguide mode, but rather a linearized transmon with its intrinsic negative nonlinearity canceled by the quarton coupler's induced positive nonlinearity. We will start by introducing the circuit for the proposed ``quartonic readout'' scheme which is designed to be experimentally feasible with the introduction of normal-metals to break unwanted galvanic superconducting loops. We then quantize the circuit and show in an example parameter study that the proposed quartonic readout scheme's parameter requirements such as large $\chi$, good transmon resonator linearization, and sufficient transmon qubit anharmonicity can be satisfied. Lastly, we show full master equation simulations of the readout performance and discuss the preservation of qubit coherence properties as well as scalability of the proposed scheme.

\section{Results}
\subsection{Quartonic readout circuit}

\begin{figure}
    % \centering
    \includegraphics[width=0.475\textwidth]{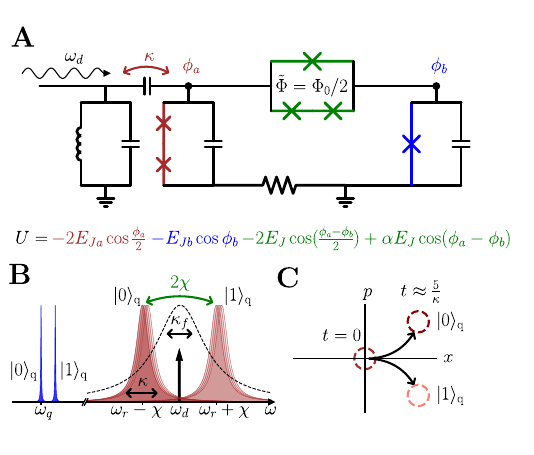}
    \caption{\textbf{Circuit diagram and readout method.} (\textbf{A}) The quarton (green) couples the linearized readout resonator (red) to the qubit (blue). A Purcell filter prevents qubit state leakage through the resonator. (\textbf{B}) Frequencies of the circuit. The readout drive at $\omega_d=\omega_r$ is between the resonator frequencies for qubit state $\ket{0}$ and $\ket{1}$. The Purcell filter (dashed) is centered at $\omega_r$. (\textbf{C}) Qubit state-dependent evolution of the resonator coherent state.}
    \label{fig: 1}
\end{figure}

We first describe the readout circuit, which uses the quarton, a purely nonlinear coupler introduced in our previous work \cite{quartonPRL}. Using a quarton (green) to couple a low-anharmonicity transmon (red) and a transmon (blue), as depicted in Fig.~1A, yields a potential in terms of nodal superconducting phase $\phi$:
\begin{equation}
\label{eq:potential_raw}
\begin{aligned}
U & =-2E_{Ja}\cos\frac{\phi_a}{2}-E_{Jb}\cos\phi_b \\
& -2E_J\cos(\frac{\phi_a-\phi_b}{2})  + \alpha E_J\cos(\phi_a-\phi_b).
\end{aligned}
\end{equation}
Note that the quarton loop is biased by half a flux quantum (see Fig.~1A), leading to the positive sign of the $\alpha E_J$ term \cite{quartonPRL}. We can Taylor expand Eq.~(\ref{eq:potential_raw}) to fourth order in $\phi$ to arrive at a more convenient form:
\begin{equation}
\label{eq:potential}
\begin{aligned}
U \approx & \frac{E_{Ja}}{2}\frac{\phi_a^2}{2!} - \frac{E_{Ja}}{8}\frac{\phi_a^4}{4!} + E_{Jb}\frac{\phi_b^2}{2!} - E_{Jb}\frac{\phi_b^4}{4!} \\
& + E_Q \frac{(\phi_a - \phi_b)^4}{4!},
\end{aligned}
\end{equation}
where $E_{Ja}, E_{Jb}$ are the Josephson energies of the Josephson junctions (JJs) in the two transmon-like modes, described by $\phi_a, \phi_b$; $E_J$ denotes the Josephson energy of each of the quarton's series JJs, and $\alpha \approx \frac{1}{2}$ sets the Josephson energy of the quarton's lone JJ to optimally cancel linear coupling. It is convenient to define $E_Q\equiv 3E_{J}/8$ to capture the effective quartic energy of the quarton potential.

We make three key observations about Eq.~(\ref{eq:potential}).  First, the $E_Q$ term will supply a positive quartic ($+\phi_i^4$) term, opposite to the $-\phi_i^4$ terms supplied by the $E_{Ji}$ terms. We can exploit this to linearize (keep only $\phi_a^2$) the $\phi_a$ mode for use as a readout resonator, while still allowing sufficiently large (now positive) self-Kerr of the $\phi_b$ mode for use as a qubit. Second, the linear coupling coupling, $\phi_a \phi_b$, from the quarton's two branches ($-2E_J$ and $\alpha E_J$ terms) cancel, leaving only nonlinear coupling terms, such as $\phi_a^2 \phi_b^2$ between the qubit and resonator, which when expanded in the Fock basis contains the resonator-qubit cross-Kerr term $a^\dagger a b^\dagger b$ needed for readout. Having cross-Kerr coupling without linear coupling prevents detrimental effects such as Purcell decay that are ubiquitous in state-of-the-art dispersive readout.
Third, in contrast to the other nonlinear couplers \cite{C-shunt-SQUID, JRM_coupler}, the quarton coupler does not supply any quadratic ($\phi_i^2$) terms to directly change the linear Josephson inductance of the modes. In a linearized analysis \cite{BBQ}, the quarton behaves as an electrical open circuit. This allows the cross-Kerr provided by the quarton coupler to scale approximately linearly with $E_Q$ \cite{quartonPRL}, enabling ultrastrong cross-Kerr coupling $2\chi/2\pi \rightarrow 1$ GHz. 

We summarize some key operating frequencies of the quartonic readout system in Fig.~1B. The qubit state $\ket{0}_q\rightarrow\ket{1}_q$ and $\ket{1}_q\rightarrow\ket{2}_q$ transitions are detuned by a positive anharmonicity (from self-Kerr) and the readout resonator is nearly linear, with linewidth $\kappa$ and cross-Kerr coupling $2\chi$ to the qubit mode. Readout is performed by probing the resonator with a tone at $\omega_d$, between the two qubit-state-dependent resonator frequencies. As in dispersive readout, this results in a coherent state evolution of the resonator mode as depicted in Fig.~1C, where within a time $t \approx 5/\kappa$ the qubit states are clearly distinguishable. Therefore, by using large $\kappa \approx 2\chi$, this enables high-fidelity readout with very short measurement time.
\begin{figure*}
    \centering
    \includegraphics[width=\textwidth]{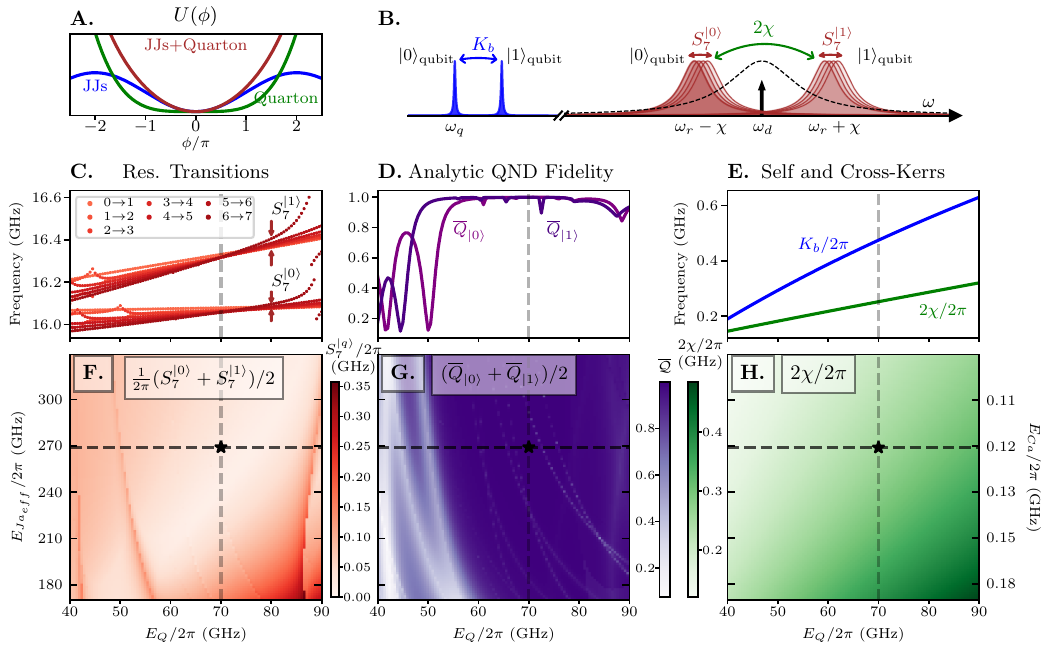}
     \caption{\textbf{Parameter sweeps for optimal parameters for readout.}  \textbf{(A)} Linearizing the resonator by cancelling the JJ and QRM $\phi^4$ terms. \textbf{(B)} Schematic eigenenergy spectrum and labeled quantities to be optimized. \textbf{(C)} Resonator transition frequencies for qubit states $|0\rangle,|1\rangle$ showing good linearization at $E_Q/2\pi=70$ GHz. \textbf{(D)} Predicted QND fidelity from analytic readout behavior as a function of $E_Q$. \textbf{(E)} Cross-Kerr between the resonator and qubit ($2\chi$) and qubit self-Kerr ($K_b$) both increasing with $E_Q$. (C-E) are respective horizontal line cuts (black dashed) in (F-H).  \textbf{(F-H)} 2D sweeps of \textbf{(F)} frequency spread, \textbf{(G)} analytic QND fidelity, and \textbf{(H)} cross-Kerr, using constant resonator frequency $\omega_a=\sqrt{8E_{Ca}E_{Ja_{eff}}}$, constant qubit parameters, and varying $E_Q$ and $E_{Ja_{eff}}/E_{Ca}$. The starred point has highly optimal parameters for readout. }
    \label{fig:2}
\end{figure*}

As shown in Fig.~1A, we also include a bandpass Purcell filter \cite{PurcellFilter2014} of width $\kappa_f = 4\kappa$ (dashed line in Fig.~1B) to prevent qubit state leakage.
In addition, to avoid unwanted flux bias in the galvanic loop formed by the quarton, the transmons, and ground, while simultaneously allowing applied flux bias in the quarton loop $\tilde\Phi = \Phi_0/2$, we propose using a highly conductive, but non-superconducting, piece of normal metal to
“break” the unwanted loop through ground (resistor in Fig.~1A). See also Appendix \ref{sec:normal_metal_bias} for more details on the effects of the normal metal resistor in the circuit, and Appendix \ref{sec:resistor_loss} for its effects on the losses in the circuit. This would allow the entire
system to be flux biased by a single current source.

\subsection{Optimal parameters for readout}
\begin{table}
    \centering
    \begin{tabular}{l c r}
        \hline
        \hline
         Res. frequency & $\omega_{10}/2\pi$ & $16.1\:\mathrm{ GHz}$\\
        Res. effective $E_J$& $E_{Ja_{eff}}/2\pi$ & $269\:\mathrm{ GHz}$\\
        Res. mode $E_C$ & $E_{Ca}/2\pi$ & $119\:\mathrm{ MHz}$\\
        Res. capacitance & $C_a$ & $163\:\mathrm{ fF}$
        \\
        %Number of JJs in Res. & $n_{Ja}$ & $2$\\
        \hline
        Qubit frequency & $\omega_{01}/2\pi$ & $7.78\:\mathrm{ GHz}$\\
        Qubit $E_J$ & $E_{Jb}/2\pi$ & $20.3\:\mathrm{ GHz}$ \\
        Qubit mode $E_C$ & $E_{Cb}/2\pi$ & $325\:\mathrm{MHz}$\\
        Qubit capacitance & $C_b$ & $59.6\:\mathrm{ fF}$ \\
        \hline
        Quartic potential& $E_Q/2\pi$& $70.0\:\mathrm{ GHz}$\\
        Quarton tilt & $2\alpha $ & $1.02$\\
        Quarton lone JJ & $\alpha E_{J}/2\pi$ & $95.2 \:\mathrm{ GHz}$\\
        Quarton series JJ & $E_{J}/2\pi$ & $186.7 \:\mathrm{ GHz}$\\
        Quarton JJ capacitance & $C_J$ & $5.2\:\mathrm{ fF}$\\
        \hline
        Res. spread for $\ket{0}_q$ & $S^{|0\rangle}_{7}/2\pi$& $21.9 \:\mathrm{ MHz}$
        \\
        Res. spread for $\ket{1}_q$& $S^{|1\rangle}_{7}/2\pi$& $15.7\:\mathrm{ MHz}$
        \\
        Qubit self-Kerr & $K_b/2\pi$ & $475\:\mathrm{ MHz}$\\
        Cross-Kerr & $2\chi/2\pi$ & $252\:\mathrm{ MHz}$\\
        \hline
        \hline
    \end{tabular}
    \caption{Summary of parameters for starred point in Fig.~2.}
    \label{tab:params}
\end{table}
Following Eq.~(\ref{eq:potential}), and adding the capacitive (kinetic) energy terms associated with Cooper pair number $\hat{n}$ operators but neglecting the normal metal resistor (see Appendix \ref{sec:resistor_loss} for its perturbative treatment), we can derive the total system Hamiltonian:
\begin{equation}\label{eq:ham}
    \begin{aligned}
        \hat{H} & =4E_{Ca}\hat n_a^2+4E_{Cb}\hat n_b^2+8E_{Cab}\hat{n}_a\hat{n}_b \\
        & -2E_{Ja}\cos\left(\frac{\hat\phi_a}{2}\right)-E_{Jb}\cos\left(\hat\phi_b\right)\\
        & -2E_{J}\cos\left(\frac{\hat\phi_a-\hat\phi_b}{2}\right) +\alpha E_{J}\cos(\hat\phi_a-\hat\phi_b).
    \end{aligned}
\end{equation}
Note that we have made the common approximation \cite{quartonPRL} to treat $n$ series JJ arrays as a single element with potential $nE_J \cos{(\phi/n)}$, see Appendix \ref{sec:internal_modes} for justification via a more detailed treatment without this approximation. We have also included a small capacitive coupling term $8E_{Cab}\hat{n}_a\hat{n}_b$ to model the Josephson junction capacitance in the quarton coupler and other stray capacitance between the resonator and qubit (assumed here to be $\sim 5\,\mathrm{ fF}$). In order to maintain net zero linear coupling, we cancel the effect of the capacitive coupling term by engineering an opposite-signed linear inductive coupling in the quarton; this is done by perturbing the value of $\alpha \approx 1/2$ or ``tilting'' \cite{quartonPRL} the quarton (see Methods).
We numerically solve (with $\hbar=1$ units) for the eigenenergies $\omega_{n_a n_b}$ of Eq.~(\ref{eq:ham}) in the Fock basis, without any rotating wave approximations (RWA), and label the eigenstates $\ket{n_a,n_b}$ by resonator ($n_a$), qubit ($n_b$) excitation number (see Methods for details). 

For fast, high-fidelity and QND qubit readout, the system should simultaneously exhibit low resonator nonlinearity, strong qubit-resonator cross-Kerr, strong qubit self-Kerr, and high predicted QNDness from analytics. We quantify resonator nonlinearity by defining the resonator frequency spread 
$S^{|q\rangle}_{n^*}:=\max_{i,j \leq n^*}|(\omega_{iq}-\omega_{(i-1)q}) - (\omega_{jq} - \omega_{(j-1)q})|$, 
where $n^*$ is the number of resonator transitions we consider, 
and $(\omega_{iq}-\omega_{(i-1)q})$ is the single photon resonator transition from eigenstates $\ket{i, q} \rightarrow \ket{i-1, q}$.
We would ideally like our linearized transmon resonator to behave as a perfect linear resonator, or $S^{|q\rangle}_{\infty} = 0$, but as shown in Fig.~2A, the linearized transmon resonator lives in an effective potential that is a sum of the transmon JJs cosine function and the quarton coupler's first-order quartic function. This results in a quadratic function potential near the bottom of the potential well, with the corresponding low energy levels being linearly spaced, but an increasingly less ideal quadratic potential and less linear energies for higher states. As such, we choose to linearize only the first $n^*=7$ transitions and will drive the resonator to a low $\bar n=2$ coherent state in subsequent readout simulations to avoid exciting the higher, nonlinear resonator states. We also compute the resonator-qubit cross-Kerr $2\chi=\omega_{11}-\omega_{01}-\omega_{10}$ and qubit self-Kerr $K_b=\omega_{02}-2\omega_{01}$. The frequency spectrum of the qubit and resonator along with the metrics $S_{7}^{|q\rangle},2\chi,K_b$ are schematically illustrated in Fig.~2B. For predicting QNDness of readout for qubit in state $\ket{k}$, we find that a concise analytic estimate $\bar{Q}_{\ket{k}}$ matches well with  numerical readout simulations. This $\bar{Q}_{\ket{k}}$ makes the realistic assumptions that the resonator quickly evolves into a steady state coherent state \cite{khezri2016} and that qubit decay into different qubit number eigenstates during readout is the main source of reduced QND fidelity (see Appendix \ref{sec:Q_bar} for details).

We now perform a parameter sweep with experimentally realistic parameters (see Table 1) to demonstrate how optimal parameters for readout can be found. We choose a transmon qubit with uncoupled frequency 7.5 GHz and a readout transmon resonator with uncoupled frequency $\omega_a/2\pi = 16$ GHz. We also use the previously defined $E_Q\equiv 3E_{J}/8$ from the quarton potential  $U_Q(\phi)\approx\frac{E_Q}{4!}\phi^4+\mathcal O(\phi^6)$ \cite{quartonPRL}. In Fig.~2F-H, we sweep both $E_Q$ and the effective resonator Josephson energy $E_{Ja_{eff}}= E_{Ja}/2$ while keeping the resonator frequency approximately constant by fixing $\omega_a\approx\sqrt{8E_{Ca}E_{Ja_{eff}}}$. At every point, we also solve for the optimal quarton tilt parameter $2\alpha$ (which is heuristically optimized as described in Methods). The sweep reveals a trade-off between resonator linearization (minimizing $S^{|q\rangle}_{7}$) and maximizing $2\chi$. 
Additionally, we see that increasing $E_{Ja_{eff}}$ of the resonator increases the $E_Q$ that minimizes $S^{|q\rangle}_{7}$ as consistent with Fig.~2A. See also Appendix \ref{sec:squeezing} and \ref{sec:Kerr_summary} where we provide a more thorough analysis and summary of the Kerr effects in the system, including the general case of more than 2 series JJs and exact treatment of coupling terms such as $(b^{\dagger\,2}+b^2)a^\dagger a$ (photon-enhanced squeezing).  The discontinuities in $S^{|q\rangle}_{7}$ are the result of avoided crossings between the eigenstates (such as between $|7,1\rangle,|5,5\rangle$ seen at the right side of Fig.~2F). It is important to operate away from such crossings, as they indicate strong hybridization of the resonator mode with the qubit mode, leading to qubit state leakage during readout and low QND fidelity (see Fig.~2G). Irregular peaks in Fig.~2F can be seen to correspond to lower QND fidelity regions in Fig 2G, where frequency collisions involving lower energy (and thus more populated in readout with $\bar{n}=2$) resonator eigenstates tend to decrease the QND fidelity more. 

In Fig.~2C-E, we take a horizontal line cut in the 2D sweep (black dashed line in Fig.~2F-H) to examine the effect of $E_Q$ alone. Fig.~2C shows the resonator transition frequencies for both qubit states and for various energy levels $i \rightarrow i + 1$, a clear signature of linearization at $E_Q/2\pi = 70$ GHz appears (grey dashed line), where resonator frequency spread for both qubit states $S^{|0\rangle}_{7}$, $S^{|1\rangle}_{7}$ is less than 25 MHz. Furthermore, Fig 2D illustrates the variations in the analytically predicted QND fidelity for both qubit states, showing that both $\bar{Q}_{\ket{0,1}} > 99\%$ at $E_Q/2\pi = 70$ GHz. In Fig.~2E, we see the expected \cite{quartonPRL} trend of both the cross-Kerr and qubit self-Kerr $K_b$ increasing with increasing $E_Q$. At $E_Q/2\pi = 70$ GHz, we find over $250\:\mathrm{ MHz}$ of cross-Kerr and over $400\:\mathrm{ MHz}$ of qubit self-Kerr. Since $\kappa\approx2\chi$, this means $S^{|q\rangle}_{7}\ll\kappa$ and we can drive the resonator to a coherent state. In summary, the point marked with a star in Fig.~2F-H is an example of a parameter set (see Table 1) that satisfies all the criteria for optimal quartonic readout. We will proceed to use this parameter set for the subsequent readout dynamics simulations. We also emphasize that the proposed quartonic readout scheme is  versatile and many other suitable parameter sets exists for other resonator and qubit frequencies. 
\begin{figure}
\includegraphics[width=0.4\textwidth]{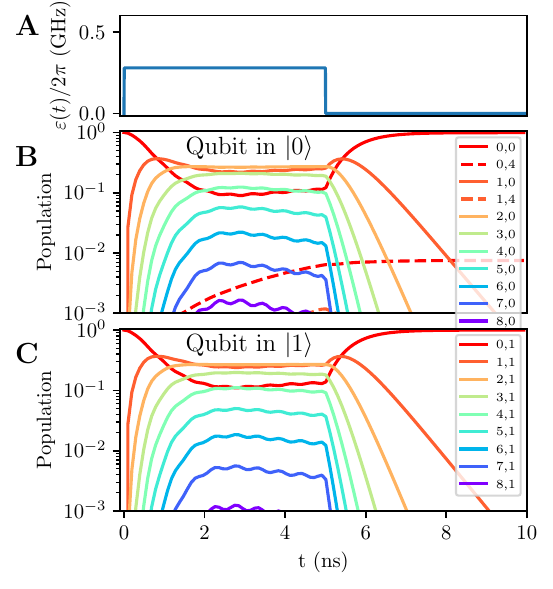}
\label{fig:QND}
\caption {\textbf{Quartonic readout simulation showing near-ideal evolution with high QNDness.} ($\textbf{A}$) Square drive pulse for simulation. 
($\textbf{B-C}$) 
Eigenstate population for readout of qubit initially in  ($\textbf{B}$)  qubit eigenstate $\ket{0}$ and ($\textbf{C}$) $\ket{1}$  with eigenstate convention $\ket{\text{resonator, qubit}}$. The state evolutions are highly QND with no significant leakage into other qubit states.}
\end{figure}
\begin{figure*}
\includegraphics[width=0.75\textwidth]{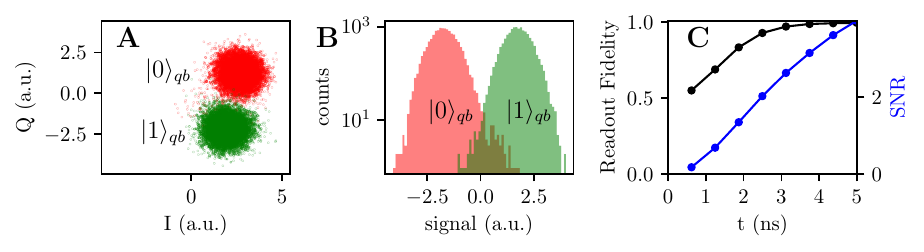}
\label{fig:readout-fidelity}
\caption {\textbf{Quartonic readout statistics showing high readout fidelity for 5 ns square readout pulse.}($\textbf{A}$) Stochastic master equation simulated time-integrated measurement results in the IQ plane; 12,800 points each represents a measurement trajectory, red (green) points are for qubit $\ket{0}$ ($\ket{1}$). ($\textbf{B}$) Histogram of measurement signal from $\textbf{A}$. ($\textbf{C}$) Readout fidelity and SNR over time (duration of square pulse). }
\end{figure*}

\subsection{Readout performance}

Using the optimized system parameters in Table \ref{tab:params}, we simulate performance with $\eta = 1, \kappa/2\pi = 300 $ MHz and a readout drive resulting in steady-state $\bar{n}=2$ photons. Key performance results are summarized in Table \ref{tab:transmon_fidelity}. Numerical full master equation simulations show near-ideal state evolution over a readout pulse (Fig.~3A) duration of 5 ns with high average QND fidelity of 99.55\% in Fig.~3 and high average readout fidelity $99.60\%$ in Fig.~4. We emphasize that we performed these simulations in the lab frame with the full JJ cosine potential and without applying rotating wave and dispersive approximations, see Methods for more details.

As shown in Fig.~3A, we apply a resonator drive $\varepsilon(t) = u(t)  [2\varepsilon_0 \cos{(\omega_d t)}] u(5-t)$, for Heaviside step function $u(t)$, which is a simple square pulse of width 5 ns followed by post-readout resonator ring-down time (5 ns). The resulting system dynamics are numerically computed with the Lindblad master equation solver in QuTiP \cite{qutip} with all dissipations treated rigorously following Ref. \cite{beaudoin2011} (see Methods for details). 
In Fig.~3B-C, we plot the time evolution of the eigenstate (for the undriven Hamiltonian) population for qubit initialized in eigenstate $\ket{0}$ (Fig.~3B) and $\ket{1}$ (Fig.~3C), respectively. 
As expected, the large $\kappa$ of quarton coupled readout resonator allows for very fast resonator response ($\approx 1$ ns) to the drive $\varepsilon(t)$.
We also observe near-ideal behavior, with the resonator reaching an equilibrium coherent state from the drive while the qubit state remains mostly unchanged. Final states at the end of the readout and ring-down periods shows very little leakage outside the starting qubit computational states, translating to a high QND fidelity of 99.13\% (99.98\%) for qubit initialized in $\ket{0}$ ($\ket{1}$). Here, we use the standard definition of QND fidelity as being the probability of the qubit remaining in state $\ket{i}$ after measuring $\ket{i}$ \cite{Sunada2022-IntrinsicPF}. The excellent QND fidelity of our proposed readout results from the non-perturbative cross-Kerr between the readout resonator and qubit, with the main parasitic coupling term being two-photon hopping of the form $\hat{a}^{\dagger2} b^2 + \text{h.c.}$, which are highly suppressed from the massive frequency detuning between the 7.73 GHz qubit and the 16.08 GHz resonator. These two-photon hopping terms are still visible in Fig.~3B-C as they cause qubit leakage from starting state $\ket{i} \rightarrow \ket{i + 2n}$ for integer $n$. 
The high QND fidelity found in numerical simulation closely matches the analytic expectation of high $\bar{\mathcal{Q}}$ in Fig.~\ref{fig:2}, justifying the construction of the analytic metric.

Simulated readout statistics are shown in Fig.~4. Here, we numerically compute measurement trajectories using the stochastic master equation \cite{jacobs2006} with heterodyne detection. We include all the same dissipations in the previous Lindblad master equation simulation, except we choose to monitor the dissipation from the relevant resonator decays for our measurement operator (see Methods). The 12,800 measurement trajectories are demodulated to extract the $I(t), Q(t)$ quadratures,
and following state-of-the-art experimental readout protocols \cite{walter2017}, we simulate the integrated $I,Q$ quadrature signals (see Fig.~4A) by time-integrating the trajectories $I(t), Q(t)$ with weighting functions $W_{Q}(t) \propto\left|\left\langle Q_{\ket{1}}(t)-Q_{\ket{0}}(t)\right\rangle\right|$ and $W_{I}(t) \propto\left|\left\langle I_{\ket{1}}(t)-I_{\ket{0}}(t)\right\rangle\right|$. Each ($I,Q$) point in Fig.~4A is obtained by time-integrating a trajectory over the max readout pulse length of 5 ns, which after an optimal axis projection gives the histograms of signal in Fig.~4B. From Fig.~4B, a readout fidelity of 99.52\%, 99.69\% (for qubit initialized in $\ket{0}, \ket{1}$, respectively) can be extracted from histogram overlaps \cite{blais2021}. 

In Fig.~4C, we simulate the readout performance as a function of pulse length and plot the resulting readout fidelity and signal-to-noise ratio (SNR). This is done by integrating the measurement trajectories over times ranging from 1-5 ns. Results here confirm that our proposed quartonic readout scheme with a high $\kappa, \chi$ readout resonator indeed results in much faster SNR and readout fidelity growth with measurement time. We note that measurement times beyond 5 ns could improve readout fidelity further, albeit at the expense of lower QND fidelity below 99\% for qubit initialized in $\ket{0}$ since leakage into qubit state $\ket{4}$ would increase to more than 1\% (see Fig.~3B). 

\begin{table}
    \centering
    \begin{tabular}{|c c c c|}
        \hline
        \hline
        Qubit initial state & Readout Time (ns) & Readout fidelity & QND-fidelity \\
        \hline
        \hline
        $\ket{0}$ & 5 & 99.52\% & 99.13\% \\
        \hline
        $\ket{1}$ & 5 & 99.69\% & 99.98\% \\
        \hline
        \hline
    \end{tabular}
    \caption{Summary of simulated transmon qubit readout performances.}
    \label{tab:transmon_fidelity}
\end{table}

\section{Discussion}

\subsection{Purcell Decay and Shot Noise Dephasing}
Ideally, any readout scheme should not significantly worsen the qubit's decoherence ($T_1, T_2$). Here, we examine how common readout-induced decoherence channels such as Purcell decay \cite{PurcellDecay} and thermal shot noise dephasing \cite{bertet2005, EQuS_thermal_photon} can be suppressed in the proposed quarton-coupler readout scheme.  

Without the Purcell filter, the readout resonator is coupled via its (normalized) charge operator $\hat{n}_0 := i(\hat{a}^\dagger - \hat{a})$ to the readout environment (or bath) with frequency dependent rate $\kappa(\omega)$, and the qubit is coupled via the tilted quarton to the readout resonator, the eigenstates transitions of the system $\ket{e_j} \rightarrow \ket{e_i}$ are effectively coupled to the bath with rate $\Gamma_{ji}=\kappa(\omega_{ji}) |\bra{e_j}\hat{n}_0\ket{e_i}|^2$ \cite{beaudoin2011}. So for qubit eigenstates $\{\ket{e_j},\ket{e_i}\} = \{\ket{1},\ket{0}\}$, the relaxation rate is enhanced by the Purcell decay rate $\Gamma_{P} \equiv \Gamma_{10}$. By virtue of quarton couplers being purely nonlinear couplers, it is possible to use a gradiometric quarton \cite{gradiometric} as the coupler and apply in-situ flux-tuning to exactly cancel all linear coupling between the qubit and readout resonator \cite{quartonPRL, kounalakis2018}, thereby making  $\Gamma_{P} = 0$.
It is worth emphasizing that quarton-coupler's zero Purcell decay with large cross-Kerr can be achieved \textit{without} Purcell filters, in contrast to state-of-the-art dispersive readout where Purcell decay cannot be avoided without Purcell filters due to the underlying linear coupling required for cross-Kerr. 

In dispersive readout, large $\Gamma_P$ is typically suppressed by adding Purcell filters \cite{PurcellFilter2010, PurcellFilter2014, IBM_Stopband_PF, IBM_NotchFilter_PF, Sunada2022-IntrinsicPF} that minimize qubit $\kappa(\omega_q) \rightarrow 0$ while keeping resonator $\kappa(\omega_r)$ unchanged. While a Purcell filter is not required for quarton-coupler readout to eliminate Purcell decay at $\kappa(\omega_q)$, it is nevertheless beneficial for suppressing unwanted decay at other eigenstate transitions (see Appendix \ref{sec:ME_sim} for details). 
Purcell filtering is highly compatible with quarton-coupler readout since transition frequencies to filter are usually many gigahertz away from the resonator frequency $\omega_r$ we wish to preserve. State-of-the-art bandpass Purcell filters with quality factors $Q_f \approx 10$ can be used, which is sufficiently low \cite{heinsoo2018} compared to the quality factor of the readout resonator $ Q_r = \omega_r / \kappa(\omega_r) \approx 50$ for the Purcell filter to operate ideally \cite{Sete2015}.

Qubit dephasing from shot noise of thermal photons in the resonator is another important source of decoherence that increases with $\kappa$ (for fixed $\chi/\kappa$) \cite{Gambetta2008, blais2021}. For an average thermal photon number $\bar{n}_t$, the shot noise dephasing rate is given by \cite{bertet2005}
\begin{equation}
    \gamma_m = \frac{\bar{n}_t(\bar{n}_t+1)(2\chi)^2}{\kappa(\omega_r)}.
\end{equation}
We can substantially reduce $\gamma_m$ in quartonic readout by leveraging both the frequency-tunability and the higher resonant frequency of quarton coupled readout resonators. 
The frequency-tunability of quarton coupled readout resonators stems from the use of a lumped-element transmon-like mode for readout, with inductance provided by JJs which can be replaced with SQUIDs that can be in-situ flux-tuned \cite{series_SQUID}. This allows for potential schemes that flux-tune the readout resonator into a very low $\kappa$ frequency band of the readout transmission line (e.g. protected by a filter) to reduce $\gamma_m$ before readout, and flux-tune the resonator back to the desired $\kappa$ frequency during readout. 
Such active flux-tuning schemes may also be replaced by a more hardware-efficient, passive scheme of shot noise protection in the form of a well-thermalized \cite{45mK,EQuS_thermal_photon} high frequency readout resonator \cite{Sunada2023}. Unlike dispersive readout, quartonic readout performance does not explicitly depend on qubit-resonator frequency detuning, so the quarton coupled readout resonator can be made very high frequency (e.g.$\approx 16$ GHz in Table 1) relative to current transmon qubits and dispersive readout resonators (typically 3-8 GHz \cite{EngineersReview}). Since $\gamma_m$ scales directly with the average thermal photon population following Bose-Einstein statistics: $\bar{n}_t(T) = [\exp{(\hbar \omega_r / (k_B T))} - 1]^{-1}$, a marginally higher resonator frequency $\omega_r = 2\pi \times 12.5 $ GHz with state-of-the-art thermalization (effective temperature $T = 45 $ mK \cite{45mK}) can have $10^2$ times lower $\bar{n}_t$ and thus $10^2$ times lower $\gamma_m$ compared to a state-of-the-art dispersive readout resonator $\omega_r = 2\pi \times 8 $ GHz, thereby nullifying the impact of $10^2$ times larger $\kappa$. In fact, with even larger $\omega_r/2\pi \gg 12.5 $ GHz easily achievable for quartonic readout, the $\gamma_m$ increase from larger $\kappa$ may be more than offset by many orders of magnitude lower $\bar{n}_t$, making the quartonic readout's $\gamma_m$ net lower than state-of-the-art dispersive readout's. 
\begin{table}
    \centering
    \begin{tabular}{|ccc|} 
        \hline
        \hline
        Mechanism & Decoherence rate $\Gamma_1$ & $T_{1} = 1/\Gamma_1$ \\ 
        \hline
        \hline
        \rule{0pt}{3ex} 
        Purcell decay & $\kappa(\omega_q) |\bra{0,0}\hat{n}_0\ket{0,1}|^2$  & - \\
        \rule{0pt}{3ex} 
        normal metal & Eq.~(\ref{eq:resistloss}) & 113 ms \\
        \rule{0pt}{3ex} 
        quasiparticle & $\left|\bra{0}\sin \frac{\hat{\phi}}{2}\ket{1}\right|^2 \frac{8 E_{J}}{\pi \hbar} x_{\mathrm{qp}} \sqrt{\frac{2 \Delta}{\hbar \omega_{q}}}$ & 0.42 ms \\
        \rule{0pt}{3ex} 
        dielectric loss & $\frac{\hbar \omega_{q}^2}{4 E_C Q_{\text {diel }}}|\langle 0|\hat{\phi}| 1\rangle|^2\left[\operatorname{coth}\left(\frac{\hbar \omega_{q}}{2 k_B T}\right)+1\right]$ & 72 $\mu$s \\
        \rule{0pt}{3ex} 
        flux noise & $|\bra{0}\frac{\partial \hat{H}}{\partial \Phi}\ket{1}|^2 S_\Phi(\omega_q)$  & 3.9 ms \\
        \hline
        \hline
         & Dephasing rate $\Gamma_2$ & $T_2= 1/\Gamma_2$ \\
        \hline
        \hline
        \rule{0pt}{3ex} 
        thermal photon & $\frac{\bar{n}_t(\bar{n}_t+1)(2\chi)^2}{\kappa}$ & 0.51 ms \\
        flux noise & simulated &  1.8 ms\\
        \hline
        \hline
    \end{tabular}
    \caption{Summary of quarton coupled qubit decoherence properties}
    \label{tab:T1}
\end{table}
\subsection{Relaxation from Normal Metal}
Because the quarton requires precise flux biasing with $\tilde\Phi = \Phi_0/2$ without flux biasing the larger loop through ground, we have incorporated a normal metal segment, modeled as a resistor, to eliminate DC flux bias in the larger loop as depicted in Fig.~\ref{fig: 1}. Although dissipation from resistors would normally lead to large decoherence rates, the nonlinear nature of the quarton coupler suppresses the current that would pass through the normal metal. We derive (see Appendix \ref{sec:resistor_loss}) the transition rate $\Gamma_{R,b}$ for the qubit's $T_1$ relaxation to be
\begin{widetext}
\begin{equation}\label{eq:resistloss}
    \begin{gathered}
    \Gamma_{R,b}=\frac{8e^2R\omega_b}{\hbar}\Big|\langle 01|\left(\frac{C_J}{C_a+\frac{C_JC_b}{C_b+C_J}}\hat n_a+\frac{C_J}{C_b+\frac{C_JC_a}{C_a+C_J}}\hat n_b-\frac{C_J^2}{C_\Sigma}(\hat n_a-\hat n_b)\right)|00\rangle\Big|^2\\
+\frac{8e^2R}\hbar\frac{(E_{J}/\hbar)^2}{\omega_b}\Big|\langle 01|\left(\frac{\alpha E_{J}}{E_{J}}\sin\left(\hat\phi_a-\hat\phi_b\right)-\sin\left(\frac{\hat\phi_a-\hat\phi_b}{n_S}\right)\right)|00\rangle\Big|^2%\\
    \end{gathered}
\end{equation}

\end{widetext}
where $C_J$ is the quarton's intrinsic capacitance and $R$ is the resistance of a normal metal segment. With normal metal junctions of $10\, \text{\textmu}\mathalpha{\Omega}$ resistance, we estimate that the qubit's $T_1$ lifetime due to resistive dissipation is $1/\Gamma_{R,b}\approx 0.11\,\text{s}$ at the operating point depicted in Fig.~\ref{fig:2}. We can calculate a similar dissipative loss for the resonator mode too, resulting in $1/\Gamma_{R,a}\approx 0.49\,\text{s}$. 
\subsection{Relaxation from Quasiparticles}
Non-equilibrium quasiparticles are a well-known source of loss for superconducting qubits \cite{blais2021, EngineersReview}. Quasiparticle decay of a superconducting qubit generally follows\cite{catelani2011} 

\begin{equation}
\Gamma_{qp} = \left|\bra{0}\sin \frac{\hat{\phi}}{2}\ket{1}\right|^2 \frac{8 E_{J}}{\pi \hbar} x_{\mathrm{qp}} \sqrt{\frac{2 \Delta}{\hbar \omega_{q}}},
\label{eq:T_qp}
\end{equation}
where $\omega_q$ is the qubit frequency, and $E_J$ and $\hat{\phi}$ are the Josephson energy and the phase operator for the junction that the quasiparticles tunnel through. Applying Eq.~(\ref{eq:T_qp}) above to each junction in our circuit, and using state-of-the-art values of quasiparticle density $x_{qp} = 5\times10^{-9}$ \cite{ms_coherence} and superconducting gap of aluminum $\Delta/2\pi = 82$ GHz \cite{blais2021}, we estimate $T_{1,qp} = 1/\Gamma_{qp}$ to be approximately 0.42 ms for the transmon qubit parameters we used for readout simulation. These values are about an order of magnitude worse than the inherent qubit $T_{1,qp}$ without quarton coupling as a direct consequence of the quarton adding additional high Josephson energy $\sim E_Q$ Josephson junctions to the qubit. We chose parameters of high $E_Q$ ($\sim 5$ times higher than intrisinc transmon $E_J$) because we opted for very high cross-Kerr $\chi \propto E_Q$ \cite{quartonPRL} to demonstrate ultrafast readout. Quasiparticle loss mitigation techniques may be applied to suppress $T_{1,qp}$ without sacrificing readout speed. Examples of mitigation techniques include quasiparticle traps \cite{trap_qp} and shielding \cite{shielding_qp}.  

\subsection{Decoherence from Flux Noise}
Flux noise is a well-known decoherence channel with ``quasi-universal'' noise power spectrum \cite{EngineersReview}: $S_{\Phi}(\omega)=A_{\Phi}^2\left(\frac{2 \pi \times 1 \mathrm{~Hz}}{\omega}\right)^{\gamma_{\Phi}}$ with $\gamma_{\Phi} \approx 0.8-1.0$ and $A_{\Phi}^2 \approx\left(1 \: \mu \Phi_0\right)^2 / \mathrm{Hz}$. Qubit relaxation from flux noise follows \cite{EngineersReview}:
\begin{equation}\label{eq:flux_noise}
\Gamma_{1, \Phi}=\left|\bra{0}\frac{\partial H}{\partial \Phi}\ket{1}\right|^2 S_{\Phi}\left(\omega_q\right)
\end{equation}
with a non-vanishing matrix element $\left|\bra{0}\frac{\partial H}{\partial \Phi}\ket{1}\right|^2  \approx E_Q^2 \phi_{zpf,b}^2$ for our quarton coupling Hamiltonian. We assume that flux noise through the quarton loop and the ground loop are independent, and apply Eq.~(\ref{eq:flux_noise}) to each loop, adding the resulting $\Gamma_{1, \Phi}$ together to get a $T_{1, \Phi}$ of 3.9 ms for the parameters in Table 1 (see Appendix \ref{sec:flux_noise} for details). We note that the normal metal only prevents DC bias of the ground loop, not AC flux noise.

We estimate pure dephasing from flux noise by numerically generating an ensemble of flux noise time series with power spectrum given by $S_{\Phi}(\omega)$. For each time series, we compute the qubit frequency as a function of time and integrate this to compute the resulting qubit dephasing. Again assuming noise in each loop is independent and fitting an exponential decay to the dephasing from each loop, we obtain a combined $T_{2, \Phi}$ of 1.8 ms (see Appendix \ref{sec:flux_noise} for details).

\subsection{Relaxation from Dielectric Loss}
Superconducting qubits suffer relaxation from dielectric loss at rate \cite{fluxonium_blueprint}:
\begin{equation}
  \Gamma_{\text {diel }}=\frac{\hbar \omega_{q}^2}{4 E_C Q_{\text {diel }}}|\langle 0|\hat{\phi}| 1\rangle|^2\left[\operatorname{coth}\left(\frac{\hbar \omega_{q}}{2 k_B T}\right)+1\right].  
\end{equation}
Using experimental $Q_{\mathrm{diel}}=7 \times 10^6$ value \cite{Q_transmon_7M} with our quarton coupled qubit frequencies and matrix elements, we compute $T_{1, \text{diel}} = 1/\Gamma_{\text {diel }} \approx 72$ 
 $\mu$s for the transmon qubit in Table 1. Unlike all previous decoherence calculations, dielectric loss does not depend explicitly on the unusually large parameters such as $E_Q$ or $\kappa$ in quartonic readout, so as expected, $T_{1, \text{diel}}$ is essentially unaffected by the quarton.
\subsection{Future directions}
The high fidelities simulated using only a simple square pulse demonstrate the robustness of our proposed scheme and leave ample room for further improvements. 
For instance, existing dispersive readout optimization techniques such as pulse shaping \cite{CLEAR, ML-PulseShaping} or qubit  shelving \cite{noTWPA} should be readily applicable to our quartonic readout scheme for some constant factor improvements in measurement time. 
Another avenue of improvement could be to leverage the inherent nonlinearity in our transmon-based readout resonator for bifurcation \cite{Sunada2023} to enhance readout performance.
More comprehensive parameter sweep and optimization could also show operation points with small $S^{|q\rangle}_{n^*}$ for larger $n^*$, which allows for more photons $\bar{n}$ to be used in readout for even higher SNR. However, higher $\bar{n}$ is also known to cause deleterious effects such as reduced qubit $T_1$ \cite{hanai2021} which are not included in our model; it is worth investigating if this and other large $\bar{n}$ effects (known from dispersive readout) like chaos-induced quantum demolition \cite{Chaos-DispersiveReadout} are relevant for large $\bar{n}$ quartonic readout.

We also emphasize that while our results here use the much higher available SNR of quartonic readout to reduce measurement time, it is conceivable that important use cases such as large-scale error correction setups \cite{GoogleSurfaceCode22} could prioritize hardware-efficiency over fast feedback time. It may therefore be advantageous to instead use the higher available SNR on tolerating lower measurement quantum efficiency $\eta$ in the hardware setup, thereby removing the need for quantum-limited amplifiers \cite{JRM_JPA, TWPAScience, FloquetTWPA} and their accompanying impedance-matching circulators or isolators \cite{TWPAScience,FloquetTWPA}. Since these quantum-limited amplifiers typically improve $\eta$ by about 10 times \cite{EngineersReview}, and quartonic readout improves readout SNR per unit time by more than 10 times, we envision state-of-the-art 50 ns or less measurement time without quantum-limited amplifiers \cite{noTWPA} should be feasible, representing a drastic reduction in measurement hardware complexity. However, this would require further optimization for an operating point with much lower qubit leakage rate such that quartonic readout with a much longer (order 50 ns) duration does not result in significantly lower QND fidelity.  

\subsection{Feasiblity and scalability}
All proposed superconducting and normal metal circuit necessary for quartonic readout should be compatible with standard microfabrication. 
In particular, the normal metal allows for a simple flux bias scheme wherein all quarton couplers on a chip are designed with identical geometric loops that can be simultaneously flux biased by a single coil providing an uniform flux \cite{FeiYan}. Our assumed $10\, \text{\textmu}\mathalpha{\Omega}$ resistance should be achievable by using thin strips of platinum. At cryogenic temperatures, a platinum resistivity of $9.6\times 10^{-9} \, \mathalpha{\Omega} \, \text{cm}$ should be achievable \cite{pt_resistivity}. Depositing 300 nm of Pt to form a resistive segment $100\, \text{\textmu m}$ wide and $2\, \text{\textmu m}$ would then provide a normal metal segment with a resistance around $6.4\, \text{\textmu}\mathalpha{\Omega}$.

Microwave measurement hardware compatible with higher frequency measurements have also been developed \cite{JennMM23}.
Since higher frequencies are typically less utilized, the ability to operate readout resonators at much higher frequency due to the quarton coupler's non-perturbative cross-Kerr not being explicitly frequency dependent may be an important practical advantage compared to state-of-the-art dispersive readout which suffers from frequency crowding \cite{frequency-crowding}.

For important use cases such as large-scale error correction setups \cite{GoogleSurfaceCode22}, it is necessary to have frequency-multiplexed readout for hardware efficiency. This seems difficult at first glance, as resonators typically must be spaced several line-widths apart to avoid cross-talk \cite{frequency-crowding,heinsoo2018} which for large $\kappa/2\pi=300 $ MHz can strain the few gigahertz bandwidth of analog electronics \cite{QICK}. However, it has been shown in \cite{heinsoo2018} that the necessary frequency spacing between each resonator can be significantly reduced with individual Purcell filters, so this may provide a path to engineering multiplexed ultrafast quartonic readout of many qubits in the future. Alternatively, by sacrificing some readout speed, $\kappa/2\pi=100-200$ MHz designs can easily accommodate 10 multiplexed qubit readout channels with existing hardware \cite{QICK}, while still operating in a parameter regime that is difficult or impossible to reach for dispersive readouts with $2\chi \approx \kappa \lessapprox 2\pi \times 10 $ MHz. Frequency multiplexing would also be much easier if one opted to use the extra SNR from quartonic readout towards lowering $\eta$ rather than lowering measurement time, as this makes measurement pulses much longer in time and thus narrower in frequency, thereby reducing cross-talk. 

\subsection{Takeaways}
In summary, we present an experimentally-feasible, quarton coupler-based scheme for ultrafast superconducting qubit readout. Simulation results show that only 5 ns is needed to reach readout and QND fidelity above $99\%$, representing significant potential improvements to state-of-the-art experiments that require 40 ns \cite{Sunada2023} and could thus lead to much faster feedback schemes such as quantum error correction protocols \cite{GoogleSurfaceCode22, BeyondBreakEven23}. 
In addition to the immediate benefits in readout speed, the ability of the quartonic readout scheme to operate with low readout photon number ($\bar{n}=2$) and huge qubit-resonator detuning could alleviate some practical readout-related issues such as chaos \cite{Chaos-DispersiveReadout}, thermal photon dephasing \cite{Gambetta2008}, and frequency crowding \cite{frequency-crowding}. Unlike many existing directions of superconducting qubit readout improvements that focus on optimizing fundamentally constrained parameters like the quantum efficiency $\eta \leq 1$ \cite{JRM_JPA, TWPAScience, FloquetTWPA} and dispersive cross-Kerr and environmental coupling $2\chi \approx \kappa \lessapprox 10$ MHz \cite{walter2017,Sunada2022-IntrinsicPF, Sunada2023}, our work suggests that better non-perturbative nonlinear couplers can overcome traditional design constraints to reach new regimes of much larger $\chi, \kappa$ and thus
significantly improved readout speed.

\section{Methods}\label{section:methods}
\subsection{Fock Basis Treatments}

With a non-perturbative coupler between qubit and resonator modes, the circuit's eigenstates may differ significantly from a naive set of basis states. We represent the circuit eigenstates with the Fock basis, i.e. the Hilbert space is spanned by tensor product of Fock states in the resonator and qubit mode subspace.
Using the Fock basis is convenient because its basis states can be chosen to be very close to the eigenstates. This is essential for utilizing heuristics such as minimizing linear coupling terms ($a^\dagger b$), but maintaining the cross-Kerr terms $a^\dagger a b^\dagger b$.

The first step is writing the Hamiltonian of Eq.~(\ref{eq:ham}) in some abstract Fock basis with the canonical transformations $\hat\phi_j\to\phi_{zpf_j}(a_j^\dagger+a_j),\hat n_j\to in_{zpf_j}(a_j^\dagger-a_j)$, then normal ordering and separating out all the coupling terms to obtain a Hamiltonian of the form 
\begin{equation}
    H=H_a+H_b+H_{coup}
    \label{eq:H_j}
\end{equation}
where each $H_j\in\{H_a,H_b\}$ has all terms that only contain the operators $a_j,a^\dagger_j$, and each term in $H_{coup}$ contains operators from both modes. Normal ordering is important as this allows the process to be analytic and the coupling term $H_{coup}$ to be as simplified as possible.

To select an accurate basis to represent each mode $j$, we need to optimize our numerical choices of $\phi_{zpf_j}=\frac1{2n_{zpf_j}}$.  We do this by having the Fock basis states of mode $j$, $\{|0_j\rangle,|1_j\rangle,\dots|N_j\rangle\}$, to be as close as possible to the %
eigenstates of $H_j$, $\{|e_{0}^{(j)}\rangle, \dots |e_{N}^{(j)}\rangle\}$. Specifically, we aim to maximize the quantity
\begin{equation*}
    \sum_{k=0}^{N_j}|\langle k_j|e_{k}^{(j)}\rangle|^2
\end{equation*}
where we choose $N_j$ depending on the number of energy transitions we are interested in, which is usually up to $10$. Since each bare Hamiltonian $H_j$ includes terms with $\phi_{zpf_j}$ from the other mode, the optimization of $\phi_{zpf_j}$ for each basis isn't completely independent. However, using the procedures and heuristics explained in Appendix \ref{sec:Fock_basis}, we are able to consistently achieve overlap probabilities of over $98\%$ between the Fock basis states and the eigenstates of $H_j$ for the first 10 energy levels in each mode. This is usually sufficient for labeling the relevant eigenenergies and our relevant heuristics.

We then label the eigenstates of $H$ by iterating through each of the Fock basis states $|n_a,n_b\rangle=|n_a\rangle\otimes |n_b\rangle$ and identifying the eigenstate  $|\lambda\rangle$ (of $H$) with greatest overlap $|\langle n_a,n_b|\lambda\rangle|^2$. Then $\ket{\lambda}$ is labeled as the eigenstate version of $\ket{n_a,n_b}$. When the terms in $H_{coup}$ are not causing near-degeneracies of $H_j$ eigenstates, the labeling choices are clear due to high (over $98\%$) overlap of $H_j$ eigenstates with the basis states. As described in the main text, degeneracies between resonator and qubit excitations can lead to higher leakage rates and lower QND fidelity, so we aim to ensure high overlaps between $H_j$ (bare) eigenstates and $H$ (dressed) eigenstates.

\subsection{Quarton Tilting}

To minimize Purcell decay and unwanted mixing of the qubit and resonator modes, a main function of the quarton coupler is to have strong nonlinear (cross-Kerr) coupling without linear coupling. However, even a purely quartic $(\hat\phi_a-\hat\phi_b)^4$ coupling term, when expressed in the Fock basis with normal ordering, introduces some weak Jaynes-Cummings $a^\dagger b+ab^\dagger$ coupling terms, exactly analogous to a linear coupling term $\hat{\phi}_a \hat{\phi}_b$. This, along with unavoidable linear capacitive coupling terms $\hat{n}_a \hat{n}_b$, motivates us to modify the quarton potential by adjusting the tilt $t:=2\alpha$ to minimize the linear coupling \cite{quartonPRL}. With a Fock basis close to the bare ($H_j$ in Eq.~(\ref{eq:H_j})) eigenstates, we can tilt the quarton to minimize the quantity
\begin{equation*}
    |(\langle 0_a|\otimes \langle 1_b|)H(|1_a\rangle\otimes |0_b\rangle|^2
\end{equation*}
where $H$ is the full circuit Hamiltonian and $|n_j\rangle$ is the $n$-photon Fock state in mode $j$. This can be done by sweeping values of $t$, where we reoptimize the basis as described previously and calculate the linear coupling for each $t$. 

\subsection{Master equation simulation}
By choosing the Lindblad master equation formalism, we implicitly make the standard approximations \cite{GardinerZoller} in its derivation (e.g. Markovian and Born). We discuss the validity of this in Appendix \ref{sec:approx_in_ME_sim}.
We follow \cite{beaudoin2011} in constructing the Lindblad dissipators in the master equation by assuming a zero-temperature bath and allowing only high to low energy eigenstate transitions; it is also necessary to group transitions close ($\leq \kappa$) in frequency to the same dissipator as these transitions have correlated coupling to the same bath mode. 
In stochastic master equation simulation, the same Lindblad dissipators are used and the monitored stochastic operator is set to include only the resonator transitions for the first few qubit states. For computational efficiency, when simulating the quartonic readout system, we truncate the otherwise huge Fock-basis Hilbert space and keep only the low photon number subspaces with a threshold determined iteratively via convergence of results. For stochastic master equation simulations, 12,800 trajectories are used with 200 simulation substeps for a measurement generating time step of $1/(5\omega_d)$. 
See Appendix \ref{sec:ME_sim} for a more detailed presentation.

\section{Acknowledgment}
The authors thank Terry Orlando, Max Hays, Daniel Sank, David Toyli, Jens Koch, Aashish Clerk, Kyle Serniak, William Oliver for fruitful discussions and insightful comments; the authors would also like to thank David Rower for sample flux noise simulation code. 

This research was supported in part by the Army Research Office under Award No. W911NF-23-1-0045 and the AWS Center for Quantum Computing. Y.Y. acknowledges support from the IBM PhD Fellowship and the NSERC Postgraduate Scholarship. J.B.K acknowledges support from the Alan L. McWhorter (1955) Fellowship. S.C. acknowledges support from the MIT Undergraduate Research Opportunities Program.

\appendix

\section{\label{sec:Q_bar}Analytic QND fidelity estimate}

Here we provide a detailed construction of the analytic QND fidelity estimate used in the main text. 

We begin by approximating the system steady-state during readout as $\ket{\alpha,k}$ where the resonator is in a coherent state $\ket{\alpha}$ and the qubit is in its original number state $\ket{k}$. This is valid both in dispersive readout \cite{khezri2016} and in our quartonic readout for sufficiently good readout resonator linearization for the given drive and $\kappa$, as demonstrated in the main text. We treat qubit leakage as a small perturbation to this steady state, caused by some weak decay following Fermi's Golden Rule that takes $\ket{\alpha,k} \rightarrow \ket{n,q}$ where $\ket{n,q}$ is some eigenstate with a different qubit state $q \neq k$. We focus only on the leakage caused by incoherent decay and neglect leakage caused by coherent drive because the leakage transitions are typically far off-resonant from our readout drive frequency, so the coherent drive induced leakage is negligible compared to incoherent decay. It is convenient now to introduce a decay-induced transition matrix:
\begin{equation}
\begin{aligned}
D & = \sum_{i < j} \ket{e_i}\bra{e_j} \times  \sqrt{\kappa_{\text{eff},ij}} \\
& = \sum_{i < j} \ket{e_i}\bra{e_j} \times |\bra{e_i} \hat{n}_0 \ket{e_j}| \sqrt{\kappa(\omega_{ij}) f(\omega_{ij}})
\end{aligned}
\end{equation}
which maps any energy eigenstate $\ket{e_j}$ to lower energy eigenstates $\ket{e_i}$ (here $i,j$ are indices sorted by low to high energy) with an effective rate $\kappa_{\text{eff},ij}$. The effective rate is weighted by the transition's normalized charge ($\hat{n}_0$) coupling to the bath, the bath density of state at the transition's frequency $\omega_{ij}$, and (optionally) the filtering function $f(\omega)$ of the Purcell filter. This makes the usual assumption that the bath is at zero temperature and so the system can only lose energy to the bath. We also use our eigenstate labelling result to assign each eigenstate to some unique resonator, qubit state $\ket{e} = \ket{n,q}$.

Using the transition matrix $D$ and ignoring second-order processes like reverse leakage back to original qubit state $\ket{q\neq k} \rightarrow \ket{k}$, we can estimate the first-order leakage rate from the readout steady state $\ket{\alpha,k}$ as:
\begin{equation}
\Gamma_{\ket{k}} = \sum_{n, q\neq k} |\bra{n,q} D \ket{\alpha,k}|^2
\end{equation}
which estimates the leakage rate as the sum of decay rates to eigenstates that do not preserve qubit state ($q\neq k$). 
Finally, for readout times $\Delta t$ and ignoring transient response, we can estimate the analytic QND fidelity ($
\bar{\mathcal{Q}}$) from leakage:
\begin{equation}
\begin{aligned}
\bar{\mathcal{Q}}_{\ket{k}} & = \exp{(-\Delta t \times \Gamma_{\ket{k}})} \\
& = \exp{(-\Delta t \sum_{n, q\neq k} |\bra{n,q} D \ket{\alpha,k}|^2)}
\end{aligned}
\end{equation}

In summary, $\bar{\mathcal{Q}}_{\ket{k}}$ is an analytic estimate of the readout QND fidelity ($\mathcal{Q}_{\ket{k}}$) that accounts for leakage of population into different qubit number eigenstates over the course of a short readout time $\Delta t$ by an ideal readout state $\ket{\alpha,k}$. 

\section{\label{sec:normal_metal_bias}Flux bias with normal metal}
\begin{figure}
    \centering
    \includegraphics[width=3in]{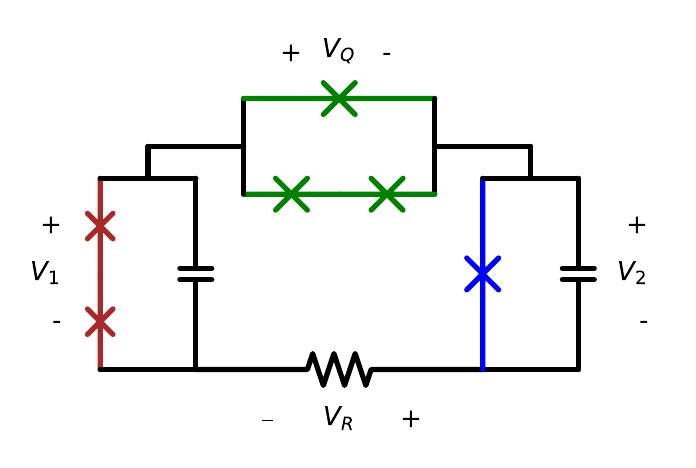}
    \caption{Voltage in circuit with resistor.}
    \label{fig:voltage}
\end{figure}
Here we show classically that the normal metal (resistor) in the circuit (Fig.~\ref{fig:voltage}) can remove DC flux bias through
the outermost loop while maintaining ideal superconducting phase behavior elsewhere. We begin with Faraday's Law for the voltages in the circuit, which in the absence of time-varying flux satisfies Kirchoff's voltage law:
\begin{equation}
V_Q + V_2 + V_R - V_1 = 0
\label{eq:Kirchoff_Voltage}
\end{equation}
In the DC steady state, there cannot be persistent current in the outermost loop or we would have energy dissipated in the resistor. This implies (in DC):
\begin{equation}
\begin{aligned}
V_R &= 0 \\
\phi_{1,2,Q} &= 0
\end{aligned}
\label{eq:DC_ss}
\end{equation}
where $\phi_{1,2,Q}$ is the superconducting phase across each of the JJs. Now we make the reasonable assumption that the resistor's effect on AC oscillation (the gigahertz modes of interest) is perturbatively small (see section ``Loss from normal metals'' for a fully quantum-mechanical calculation that makes arguments here rigorous). This can be intuitively understood from the eigenmodes of the circuit: Because the quarton is linearly an open circuit, the eigenmodes are very close to just the two individual transmons modes which has current oscillations contained within the parallel JJ and capacitor circuit and does not leak out to ground where the resistor is. So we can again take $V_R \approx 0$, then by applying the Josephson relation $\frac{\partial \phi}{\partial t} = \frac{1}{\phi_0} V(t)$ to Eq.~(\ref{eq:Kirchoff_Voltage}), we get:
\begin{equation}
\frac{\partial }{\partial t} (\phi_Q + \phi_2 - \phi_1) = 0
\end{equation}
It is reasonable to assume that the circuit starts in $t = -\infty$ in the DC steady state, then we can use Eq.~(\ref{eq:DC_ss}) as initial conditions to get:
\begin{equation}
\phi_Q + \phi_2 - \phi_1 = 0
\label{eq:flux_block}
\end{equation}
which is the ideal phase sum in the outermost loop.

Note that the only assumption used to deriving the DC steady state of Eq.~(\ref{eq:DC_ss}) was the absence of time-varying flux, implying that the circuit steady state is not sensitive to any DC applied flux ($\phi_{\text{applied}}$) in the outermost loop. Eq.~(\ref{eq:flux_block}) then followed, showing that the outermost loop in general does not follow the usual phase constraint of flux biased superconducting loops ($\sum_i \phi_i = -\phi_{\text{applied}}$). This makes intuitive sense, as the DC flux biasing of any superconducting loop is a result of persistent current flowing in the dissipationless loop. It follows that superconducting (resistor-free) loops in the circuit, such as the quarton loop, can still be DC flux biased the usual way, unaffected by the resistor which only affects the outermost loop that it is a part of. 

We emphasize that the derivation here holds only in the steady-state limit of time-independent (DC) flux bias, one would of course expect transient current through the resistor when the flux bias is changed. For typical values of circuit inductance ($L = 20 \text{ nH}$) and an expected resistance of normal metal ($R = 10\, \text{\textmu}\mathalpha{\Omega}$), we expect transient damping with a time constant of $L/R = 2 \text{ ms}$. This large time constant implies that the normal metal does not protect against high-frequency flux noise, so it would not protect qubit $T_1$ against flux noise induced relaxation (see flux noise section for details).

\section{\label{sec:internal_modes}Full circuit Hamiltonian (without series JJ approximation)}
\begin{figure}
    \centering
    \includegraphics[width=0.5\textwidth]{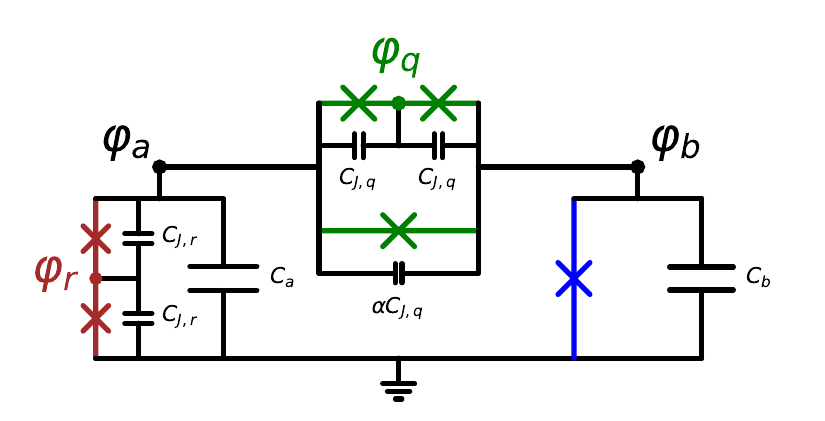}
    \caption{Labeled nodes for full circuit Hamiltonian derivation (including internal modes), assuming half flux quantum bias in the quarton loop and none elsewhere.}
    \label{fig:internal_modes}
\end{figure}

Here we show that it can be safe to ignore the internal modes (also known as ``collective modes'' in fluxonium literature \cite{collective_modes_fluxonium}) of the series JJ chains in the circuit. As shown in Fig.~\ref{fig:internal_modes}, for the representative quartonic readout circuit proposed in the main text, there are two internal modes associated with the free nodes $\phi_q, \phi_r$ in the quarton and readout resonator, respectively. In this section, we treat the normal metal as a short (Appendix \ref{sec:normal_metal_bias} for justification).

The potential energy of the circuit can be exactly expressed in terms of the nodes $(\phi_a, \phi_b, \phi_q, \phi_r)$, assuming half flux quantum bias in the quarton loop:
\begin{equation}
\begin{aligned}
U & = -E_{Ja} [\cos{(\phi_a - \phi_r)} + \cos{(\phi_r)}] - E_{Jb} \cos{\phi_b} \\
& - E_J [-\alpha \cos{(\phi_a - \phi_b)} + \cos{(\phi_a - \phi_q)} + \cos{(\phi_q - \phi_b)}]
\end{aligned}
\end{equation}
Using trig identities, we can re-write this as:
\begin{equation}
\begin{aligned}
U & = -2E_{Ja}\cos{\frac{\phi_a}{2}}\cos{\frac{\phi_a - 2\phi_r}{2}} - E_{Jb} \cos{\phi_b} \\
& - E_J [-\alpha \cos{(\phi_a - \phi_b)} \\
& + 2\cos{(\frac{\phi_a - \phi_b}{2})} \cos{(\frac{\phi_a - 2\phi_q + \phi_b}{2})} ]
\end{aligned}
\end{equation}
which is very close to the simplified form we used in the main text:
\begin{equation}
\begin{aligned}
U_{\text{eff}} & = -2E_{Ja}\cos{\frac{\phi_a}{2}} - E_{Jb} \cos{\phi_b} \\
& - E_J [-\alpha \cos{(\phi_a - \phi_b)} + 2\cos{(\frac{\phi_a - \phi_b}{2})} ]
\end{aligned}
\end{equation}
This motivates us to define a change of variable,
\begin{equation}
\begin{aligned}
\tilde{\phi}_r &= \frac{\phi_a - 2\phi_r}{2} \\
\tilde{\phi}_q &= \frac{\phi_a - 2\phi_q + \phi_b}{2}
\end{aligned}
\end{equation}
which together with the unchanged $\phi_a, \phi_b$ defines the transformation:
\begin{equation}
\begin{aligned}
\vec{\phi^{\prime}} & =W \vec{\phi} \\
{\left[\begin{array}{l}
\phi_a \\
\phi_b \\
\tilde{\phi}_r \\
\tilde{\phi}_q
\end{array}\right] } & =\left[\begin{array}{cccc}
1 & 0 & 0 & 0 \\
0 & 1 & 0 & 0 \\
1 / 2 & 0 & -1 & 0 \\
1 / 2 & 1 / 2 & 0 & -1
\end{array}\right]\left[\begin{array}{l}
\phi_a \\
\phi_b \\
\phi_r \\
\phi_q
\end{array}\right]
\end{aligned}
\end{equation}
This transformation $W$ is not unitary, so in order to maintain the canonical commutation relations between all superconducting phase $\hat{\phi}$ and Cooper pair number $\hat{n}$ operators:
\begin{equation}
\left[\phi_i, \phi_j\right]=0,\left[n_i, n_j\right]=0,\left[\phi_i, n_j\right]=i\delta_{i j}
\end{equation}
we must also transform the $\vec{n}$ by \cite{ding2021}: 
\begin{equation}
\begin{aligned}
\vec{n^{\prime}} & =(W^\top)^{-1} \vec{n} \\
{\left[\begin{array}{l}
\tilde{n}_a \\
\tilde{n}_b \\
\tilde{n}_r \\
\tilde{n}_q
\end{array}\right] } & =\left[\begin{array}{cccc}
1 & 0 & 1 / 2 & 1 / 2 \\
0 & 1 & 0 & 1 / 2 \\
0 & 0 & -1 & 0 \\
0 & 0 & 0 & -1
\end{array}\right]\left[\begin{array}{l}
n_a \\
n_b \\
n_r \\
n_q
\end{array}\right]
\end{aligned}
\end{equation}
This transforms the capacitive energy of the circuit via:
\begin{equation}
\begin{aligned}
T & =\frac{4 e^2}{2} \vec{n}^{\top} C^{-1} \vec{n} \\
& =\frac{4 e^2}{2} \vec{n}^{\prime \top} W C^{-1} W^{\top} \vec{n}^{\prime} \\
& =4 \vec{n}^{\prime \top}\left(\frac{e^2}{2} W C^{-1} W^{\top}\right) \vec{n}^{\prime} \\
& =4 \vec{n}^{\prime \top} \stackrel{\leftrightarrow}{E_C} \vec{n}^{\prime}
\end{aligned}
\end{equation}
where the capacitive energy matrix $\stackrel{\leftrightarrow}{E_C}$ is approximately diagonal but has small off-diagonal terms arising from finite junction capacitance of JJs. For instance, with experimentally-realistic estimates of capacitances: $\{C_{J,q}, C_{J,r}, C_a, C_b\}= \{3, 7.5, 80, 70 \} \text{ fF}$:
\begin{equation}
\stackrel{\leftrightarrow}{E_C}=\left(\begin{array}{cccc}
223.6 & 9.2 & 0 & 0 \\
9.2 & 265.7 & 0 & 0 \\
0 & 0 & 1291.3 & 0 \\
0 & 0 & 0 & 3228.4
\end{array}\right)  \text{ MHz} 
\label{eq:Ec_matrix}
\end{equation}
with energy in units of $h = 1$. This makes intuitive sense, as the junction capacitances in the quarton creates a direct (for $\alpha$ JJ) and indirect (for series JJ) path for capacitive coupling (non-zero $E_{C,12}$) between the two transmons; but the junction capacitances in the resonator does not contribute to any coupling. 

Putting everything together, the total circuit Hamiltonian with transformed variables is:
\begin{equation}
\begin{aligned}
H & = U + T \\
& = -2E_{Ja}\cos{\frac{\phi_a}{2}}\cos{\tilde{\phi}_r} - E_{Jb} \cos{\phi_b} \\
 & - E_J [-\alpha \cos{(\phi_a - \phi_b)} + 2\cos{(\frac{\phi_a - \phi_b}{2})} \cos{\tilde{\phi}_q}] \\
 &  + 4E_{C,11} \tilde{n}_a^2
 + 4E_{C,22} \tilde{n}_b^2 + 8 E_{C,12} \tilde{n}_a \tilde{n}_b \\
 & + 4E_{C,33} \tilde{n}_r^2
 + 4E_{C,44} \tilde{n}_q^2
\end{aligned}
\label{eq:full_H}
\end{equation}

It is clear from Eq.~\ref{eq:full_H} that the quarton and resonator internal modes are transmon-like, with cosine potential $-2 E_{J(a)} \cos{\phi_{q(r)}}$, capacitive energy $4 E_{C,44(33)} \tilde{n}_{q(r)}^2$, and no capacitive coupling. Their only coupling to the resonator, qubit modes $a, b$ is through nonlinear terms $\cos{(\phi_{a,b})} \cos(\phi_{r,q})$ which to lowest-order provides cross-Kerr like coupling $\phi_{a,b}^2 \phi_{r,q}^2$. However, these internal modes are extremely high frequency (high $E_C$ from low junction capacitance and high $E_J$ from large individual JJs used in the chain), e.g. $>35$ GHz, so they can be safely taken to be frozen in the ground state where their cross-Kerr interaction with the resonator and qubit modes $a,b$ can be ignored.
Furthermore, we can choose to fabricate the junctions in the array to have reasonable capacitances (e.g. about 5 fF such as in Eq.~(\ref{eq:Ec_matrix})) such that these internal modes have $E_J / E_C > 50$ so they can really be viewed as extremely high frequency transmons. This means the internal modes have steady frequencies and do not impart charge noise to the resonator and qubit modes (that they are nonlinearly coupled with). Therefore, we can safely ignore these internal modes, i.e. assume    $\cos{(\phi_{a,b})} \cos(\phi_{r,q}) \approx \cos{(\phi_{a,b})}$, which leads us to the simplified form of the potential energy used in main text. 
\begin{table*}
\begin{center}
\caption{\label{tab:SM-self-Kerr}Summary of self-Kerr sources and analytic scalings.}
\begin{tabular}{|c|c|c|}
\hline & Mode $a$ (resonator) & Mode $b$ (qubit) \\
\hline \hline self-Kerr from internal JJ & $-\frac{1}{n_{Ja}^2}E_{C, a}$ & $-\frac{1}{n_{Jb}^2}E_{C, b}$ \\
\hline self-Kerr from quarton coupler & $+\frac{E_Q}{E_{J, a}} E_{C, a}$ & $+\frac{E_Q}{E_{J, b}} E_{C, b}$ \\
\hline $\begin{array}{c}\text { self-Kerr from } a \text { 's squeezing } \\
a^2\left(b^{\dagger} b\right)+\text { h.c. } \\
\end{array}$ & 0 & $-\frac{\chi_{a b}^2}{2 \omega_b} = -\frac{E_Q^2 E_{C, a}^{1 / 2}}{2 E_{J, b} E^{3 / 2}_{J, a}} E_{C, b}$ \\
\hline $\begin{array}{c}\text { self-Kerr from } b \text { 's squeezing } \\
b^2\left(a^{\dagger} a\right)+\text { h.c. }\end{array}$ & $-\frac{\chi_{a b}^2}{2 \omega_a}=-\frac{E_Q^2 E_{C, b}^{1 / 2}}{2 E_{J, a} E^{3 / 2}_{J, b}} E_{C, a}$ & 0 \\
\hline
\end{tabular}
\end{center}
\end{table*}
\begin{table}
\begin{center}
\caption{\label{tab:SM-cross-Kerr}Summary of cross-Kerr sources and analytic scalings.}
\begin{tabular}{|l|l|}
\hline & Mode $a$ - $b$ (resonator-qubit) \\
\hline \hline cross-Kerr from quarton coupler & $+\chi_{a b}=2 E_Q \frac{\left(E_{C, a} E_{C, b}\right)^{1 / 2}}{\left(E_{J, a} E_{J, b}\right)^{1 / 2}}$ \\
\hline $\begin{array}{l}\text { cross-Kerr from } a \text {'s squeezing } \\
a^2\left(b^{\dagger} b\right) +\text { h.c. }\end{array}$ & $-\frac{\chi_{a b}^2}{2 \omega_b}$ \\
\hline $\begin{array}{l}\text { cross-Kerr from } b \text {'s squeezing } \\
b^2\left(a^{\dagger} a\right) +\text { h.c. }\end{array}$ & $-\frac{\chi_{a b}^2}{2 \omega_a}$ \\
\hline
\end{tabular}
\end{center}
\end{table}
\section{\label{sec:squeezing}Derivation of photon-enhanced squeezing}

One non-ideal quarton coupling effect is the addition of negative self-Kerrs and cross-Kerr to the qubit and resonator due to photon number dependent (correlated) squeezing terms of the form $(b^{\dagger\,2}+b^2)a^\dagger a$ and $(a^{\dagger\,2}+a^2)b^\dagger b$. These terms originate from the same Hamiltonian coupling term $\phi_a^2 \phi_b^2$ that gives the ideal cross-Kerr coupling, so they are unavoidable. We can model the effects of these terms by looking at a toy model of two harmonic oscillators coupled only one of these terms:
\begin{equation}
    H_{toy}=\omega_aa^\dagger a+\omega_b b^\dagger b+\zeta(b^{\dagger\,2}+b^2)a^\dagger a
\end{equation}
where we can assume $\frac{\zeta}{\omega_b}\ll1$ (generally true for quarton coupling).
We can perform a Schrieffer-Wolff transformation on this Hamiltonian with the unitary 
\begin{equation}
    S=\exp\left(\frac12a^\dagger a(z^*b^2-zb^{\dagger\,2})\right)
\end{equation}
and choose $z=re^{i\theta}$ to be real with $\theta=0$ for simplicity. By setting the coefficients of the off-diagonal $b^2$ terms in $\tilde H=SH_{toy}S^\dagger$ to 0, we obtain the condition
\begin{equation}
    \tanh(2ra^\dagger a)=\frac{2\zeta}{\omega_b} a^\dagger a
\end{equation}
and with $\zeta/\omega_b\ll1$, this is satisfied with $r=\zeta/\omega_b$. Then by expanding our transformed Hamiltonian $\tilde H$ to second order in $\zeta$, we have
\begin{equation}
    \begin{aligned}
        \tilde H &=\left(\omega_a-\frac{2\zeta^2}{\omega_b}\cosh(2ra^\dagger a)\right) a^\dagger a
         +\omega_bb^\dagger b\cosh(2ra^\dagger a)\\
         &+\omega_b\sinh^2(ra^\dagger a)
         -\frac{4\zeta^2}{\omega_b}a^\dagger ab^\dagger b\cosh(2ra^\dagger a)  \\
        & -\frac{2\zeta^2}{\omega_b}a^{\dagger\,2}a^2\cosh(2ra^\dagger a) 
         -\frac{4\zeta^2}{\omega_b}a^{\dagger\,2}a^2b^\dagger b\cosh(2ra^\dagger a)
    \end{aligned}
\end{equation}
and expanding to second order in $r$ and taking only the 4 wave mixing terms or lower, we have the transformed Hamiltonian
\begin{equation}
    \begin{gathered}
        \tilde H\approx\left(\omega_a-\frac{\zeta^2}{\omega_b}\right)a^\dagger a+\omega_bb^\dagger b-\frac{2\zeta^2}{\omega_b}a^\dagger ab^\dagger b-\frac{\zeta^2}{\omega_b}a^\dagger a^\dagger aa
    \end{gathered}
\end{equation}
which adds negative cross-Kerr $\frac{2\zeta^2}{\omega_b}$ and negative self-Kerr $\frac{2\zeta^2}{\omega_b}$ to mode $a$. This result holds for coupling via the other correlated squeezing term $(a^{\dagger\,2}+a^2)b^\dagger b$ also, which analogously adds negative cross-Kerr $\frac{2\zeta^2}{\omega_a}$ and negative self-Kerr $\frac{2\zeta^2}{\omega_a}$ to mode $b$. Note that the factor of 2 in the self-Kerr originates from the self-Kerr $K$ being $\frac{K}{2} a^\dagger a^\dagger aa$.

In typical quarton coupling circuits, the correlated squeezing term has magnitude $\zeta$ directly proportional to the ideal cross-Kerr $\chi$:
\begin{equation}
\zeta = \chi/2
\end{equation}
since they both originate from the $\phi_a^2 \phi_b^2$ coupling Hamiltonian. 

\section{\label{sec:Kerr_summary}Summary of Kerr effects in quartonic readout circuit}
Kerr nonlinearity relative to linear inductance is generally weakened by chaining together more JJs in series \cite{quartonPRL}. If we generalize the Hamiltonian in the main text to allow for any number of chained JJs in each mode, we can get a Hamiltonian:
\begin{equation}\label{eq:H_diff_n}
    \begin{aligned}
        \hat{H}&=4E_{Ca}\hat n_a^2+4E_{Cb}\hat n_b^2 +8E_{Cab}\hat{n}_a\hat{n}_b \\
        & -n_{Ja}E_{Ja}\cos\left(\frac{\hat\phi_a}{n_{Ja}}\right) 
         -n_{Jb}E_{Jb}\cos\left(\frac{\hat\phi_b}{n_{Jb}}\right) \\
        & +\alpha E_{J}\cos(\hat\phi_a-\hat\phi_b+\tilde{\phi})-n_SE_{J}\cos\left(\frac{\hat\phi_a-\hat\phi_b}{n_S}\right)
    \end{aligned}
\end{equation}

Tables \ref{tab:SM-self-Kerr}-\ref{tab:SM-cross-Kerr} below summarize the various sources of self- and cross-Kerr effects in this generalized quartonic readout setup, along with their expected analytic scalings. Two main causes of self- and cross-Kerr (other than inherent JJ self-Kerr) are the bare mode quarton Kerr effects derived in \cite{quartonPRL} and the correlated squeezing Kerr effects as derived above. A main goal in parameter optimization is to have the self-Kerr (and higher level nonlinearity) in the resonator be net zero while maintaining high (hundreds of megahertz) self-Kerr in the qubit and cross-Kerr between the qubit and the resonator. Choices of the number of series junctions $\{n_S, n_{Ja}, n_{Jb} \}$ generally have a significant influence on nonlinearity and therefore decisions should be made prudently to help achieve the goals in optimization. Generally speaking, larger $n_S$ simply increases $E_Q$ relative to $\alpha E_{J}$ \cite{quartonPRL}, whereas larger $n_{Ja}, n_{Jb}$ drastically decreases the intrinsic negative self-Kerr in mode $a, b$. This is most relevant for the resonator mode $a$, which we must linearize (achieve net zero self-Kerr). To that end, $n_{Ja}$ is an exceptionally valuable tuning knob, and we have opted for $n_{Ja} = 2$ in the main text for the particular combination of resonator and qubit frequency we were working with.
Other parameter ranges could certainly benefit from different $n_{Ja}$.

\section{\label{sec:Fock_basis}Details on Fock Basis Treatment}

\begin{figure}
    \centering
    \includegraphics[width=3in]{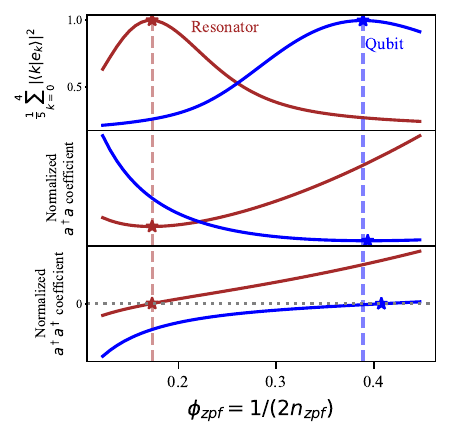}
    \caption{Different heuristics of finding $\phi_{zpf}=1/(2n_{zpf})$ values for representing eigenenergies in a nonlinear potential. The first metric finds the average overlap $\sum_{k=0}^{4}|\langle k_j|e_{k,j}\rangle|^2$ for each mode $j$. The heuristics of minimizing the coefficient of an unwanted $a^\dagger a^\dagger$ term and minimizing the first order energy $a^\dagger a$ term are displayed in comparison. The bare modes $H_a,H_b$ are calculated with $8^{th}$ order Taylor expansions of the original Hamiltonian and normal ordered with computational symbolic algebra.}
    \label{fig:zpfs}
\end{figure}

A good basis is essential for effectively simulating coupled quantum systems. As described in Methods, we split our Hamiltonian into three terms to isolate the coupling terms:
\begin{equation*}
    H=H_a+H_b+H_{coup}
\end{equation*}
and in order to accurately and efficiently represent the circuit eigenstates and their associated eigenenergies, we want to find a Fock basis that well represents the eigenstates of $H_j$
for each mode $j=a,b$. We aim to maximize
\begin{equation*}
    \sum_{k=0}^{N_j}|\langle k_j|e_{k,j}\rangle|^2
\end{equation*}
where $|e_{k,j}\rangle$ is the $k^{th}$ eigenstate of $H_j$. This translates to the general problem with finding a Fock basis for a potential $U(\phi)$ that isn't necessarily quadratic. One approach is to sweep possible values of $\phi_{zpf}=1/(2n_{zpf})$ and search for the maximum overlap as seen in Fig.~\ref{fig:zpfs}. There, we also demonstrate the performance of two analytical heuristics, first by minimizing the magnitude of the $a^\dagger a^\dagger$ term coefficient, and by minimizing the coefficient of $a^\dagger a$ terms. We may choose to implement analytical heuristics for computational efficiency.

As seen in Fig \ref{fig:zpfs}, minimizing the normal ordered $a^\dagger a$ coefficients tend to give high average overlaps between our basis and $H_j$ eigenstates. Since the Fock basis is a complete basis, the final eigenenergies of $H$ should be independent of the exact choices of the $\phi_{zpf}$ values, so the exact optimality of our $\phi_{zpf}$ values is unimportant.

Our Fock basis cannot be perfect due to the nonlinear nature of our circuit and the terms that do not preserve photon numbers (e.g. $a^2$). Additionally, one may notice that each bare Hamiltonian $H_j$ includes $\phi_{zpf_k}$ values from other modes $k\neq j$. Thus the optimization of each basis isn't completely independent. However, this ultimately doesn't affect our relevant full system eigenenergies, and with sufficiently high overlap probabilities, we can label the eigenstates without ambiguity. 

In decomposing $H$ into $H_a+H_b+H_{coup}$ to construct our bases, we normal order all the creation and annihilation operators to analytically separate terms into their respective partitions. Normal ordering is important for representing the excitations relative to the ground state, and is also important for numerical simulations, since terms such as $aa^\dagger$ in a finite Hilbert space will incorrectly map the highest Fock state to 0. This framework also helps with finding approximate analytical quantities, where we expect to see the terms $\frac{K_b}2b^\dagger b^\dagger bb\in H_b$ or $2\chi a^\dagger a b^\dagger b\in H_{coup}$. This is another reason why we aim to optimize our Fock basis, so that the creation and annihilation operators can more closely represent transitions between adjacent eigenstates.

\section{\label{sec:ME_sim}Master equation simulation}
Given full parameters to the time independent Hamiltonian $\hat{H}_0$, we use QuTiP \cite{qutip} to solve for the eigenstates and eigenenergies in the Fock basis, using a sufficiently large Hilbert space dimension for each subsystem (labelled $N_a, N_b$) such that the eigenstates of interest satisfy the commutator relations $[\hat{a}, \hat{a}^\dagger] = [\hat{b}, \hat{b}^\dagger] = 1$ numerically. In practice, this requires a large total Hilbert space size $N_a \times N_b \approx 25 \times 25$, so in order to efficiently perform the time domain master equation simulations, we change to a truncated set of eigenstates \{$\ket{e_i}$\} as our basis ($\hat{H}_{ij} = \bra{e_i}\hat{H}\ket{e_j}$). 
The truncated eigenstates are labelled by their max overlap with the bare states $\ket{ij}$, and we choose the truncated eigenstates $\ket{e_i}$ labelled with $i<i^*, j<j^*$ for imposed thresholds $i^*, j^*$.
The thresholds are iteratively increased until the dynamics (c.f. Fig.~3BC in main text) converge, with typical values of $(i^*, j^*) \approx (9, 5)$.  

Following standard treatment \cite{blais2021}, we model the resonator drive through the coupling capacitor as the time dependent operator $\hat{H}_d(t) = \varepsilon(t)\hat{n}_0 = \varepsilon(t) \times i(\hat{a}^\dagger - \hat{a})$. Then, we use the Linblad-form master equation for the system's density matrix $\rho$:
\begin{equation}
\dot{\rho}=-i\left[\hat{H}_0 + \hat{H}_d(t), \rho\right]+\sum_k \kappa_k \mathcal{D}[\hat{d}_k] \rho
\label{eq:lindblad_me}
\end{equation}
where the $k$-indexed dissipators $\mathcal{D}(\hat{d}_k) \rho ={d}_k \rho {d}_k^{\dagger}-\frac{1}{2}({d}_k^{\dagger} {d}_k \rho + \rho {d}_k^{\dagger} {d}_k)$ and their respective rates $\kappa_k$ are found by (following \cite{beaudoin2011}):
\begin{equation}
\sqrt{\kappa}_k \hat{d}_k = \sum_{i, j>i} \sqrt{\kappa_{\text{eff},ij}} \big | \bra{e_j}(\hat{a}^\dagger - \hat{a})\ket{e}_i \big| \ket{e_i} \bra{e_j}
\label{eq:kappa_k}
\end{equation}
\begin{figure}
\includegraphics[width=0.5\textwidth]{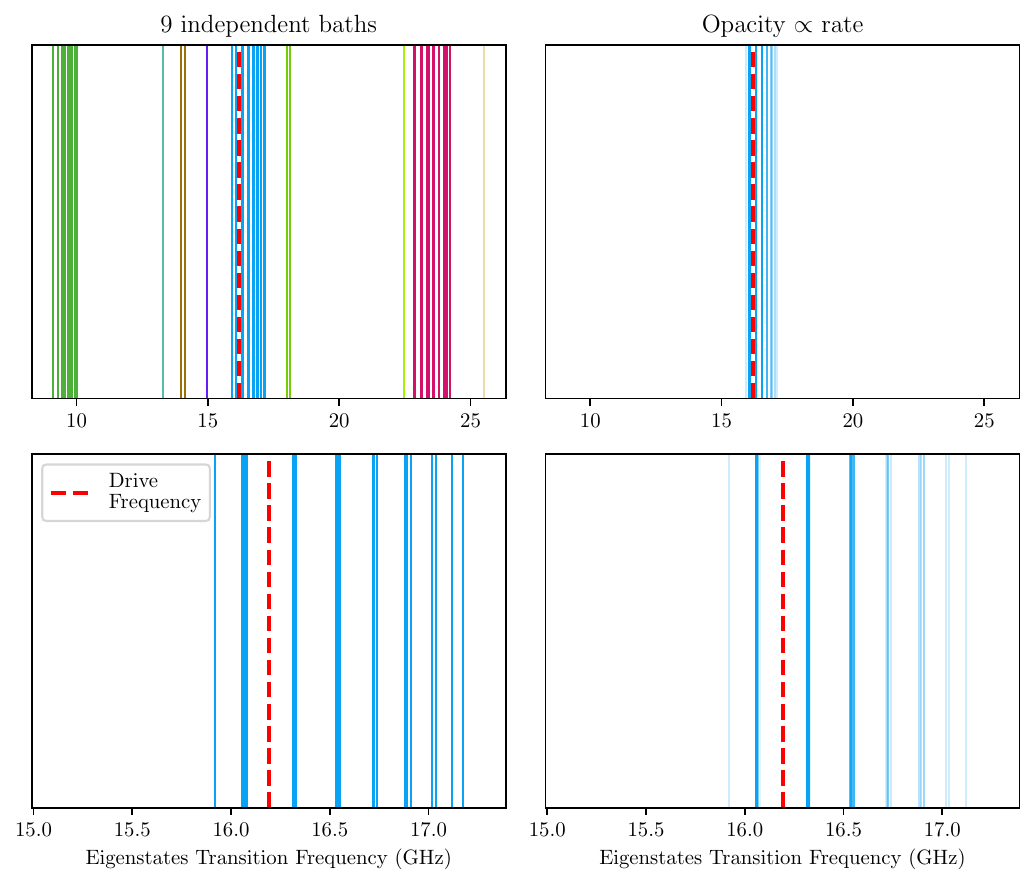}
\caption{Independent baths (different colors) found by density-based clustering algorithm (DBSCAN). Only the eigenstate transitions that most resembles resonator single photon loss have high rates (high opacity in column 2 plots). Row 2 plots are zoomed-in (around resonator and drive frequency) views of row 1 plots, showing the most important dissipation (blue) to be labelled $k^*$ and used as monitored operator in stochastic master equation.}
\label{fig:bath_modes}
\end{figure}
Note that we are explicitly choosing a zero temperature bath which can only cause transitions from high to low ($j>i$) energy eigenstates $\ket{e_j} \rightarrow \ket{e_i}$.
Furthermore, we define independent baths indexed by $k$, each coupled to the $k$'th set of eigenstate transitions $\{\omega_k\}$ that have overlapping line width \cite{beaudoin2011}. Eigenstate $j \rightarrow i$ transitions with frequencies $\omega_{ji} = \omega_j - \omega_i$ are considered to have overlapping line widths if they satisfy $|\omega_{ji} - \omega_{j'i'}| \leq c*\kappa$ (for some order unity constant $c$). In summary:
\begin{equation}
\ket{e_i} \bra{e_j} \in \hat{d}_k \text{  iff  } |\omega_{ji} - \omega_{j'i'}| \leq c \; \kappa \; \text{ for } \omega_{j'i'} \in \{\omega_k\}
\label{eq:bath_label}
\end{equation}
This is physically important as eigenstate transitions within about $\kappa$ in frequency are correlated in their coupling to the same bath mode \cite{beaudoin2011}, and independent of bath modes coupling transitions $\gg \kappa$ away. In practice, the labelling in Eq.~(\ref{eq:bath_label}) above is done via a density-based clustering algorithm (e.g. DBSCAN) on an array of all the allowed eigenstate transitions $\{\omega_{ji}\}$, to identify each of the $k$ baths. The results are shown in Fig.~\ref{fig:bath_modes}, where a threshold rate of 10 kHz was set to discard dissipations too slow to affect our $O(10)$ ns time simulations. Fig.~\ref{fig:bath_modes}'s column 2 panels repeat plots in column 1 but with opacity of lines set in proportion to the transition's $\kappa_{\text{eff},ij}$. This shows that the rates are dominated by the eigenstate transitions that most resembles resonator single photon loss, which are the transitions we monitor in readout measurement. We index this bath by $k^*$ and will use it as the monitored operator in the subsequent stochastic master equation simulation.

Note that in Eq.~(\ref{eq:kappa_k}), we use a realistic effective decay rate $\kappa_{\text{eff},ij}$ that is weighted by the coupling frequency dependence ($\kappa \propto \omega_{ji}^2$)\cite{blais2021} and the Purcell filtering response ($\kappa \propto [1+(2(\omega_{ji} - \omega_r)/\kappa_f)^2]^{-1}$) \cite{Sete2015}. The effect of Purcell filter's suppression of unwanted bath coupling far from the drive frequency is clearly visible in column 2 of Fig.~\ref{fig:bath_modes}, in practice the use of the Purcell filter improves QND and readout fidelity by a few percent, making it necessary for reaching $>\!99\%$ fidelity.
We also chose a nominal value of $c = 1$ in Eq.~(\ref{eq:bath_label}), but the results of the Linblad-form master equation simulation seem robust to a range of $c$ values we tested (around 0.5 - 2). 

Having constructed our Lindblad dissipators, we can use the Lindblad-form master equation of Eq.~(\ref{eq:lindblad_me}) to find the average dynamics of the resonator-qubit during readout, producing plots like Fig.~3 in the main text. This is sufficient for determining properties such as QND fidelity. 
However, to rigorously obtain readout fidelity, we instead simulate measurement trajectories using the stochastic master equation \cite{qutip}: 
\begin{equation}
d \rho(t)=d_1 \rho d t+ d_{2} \rho d W
\end{equation}
\begin{equation}
d_1\rho=
-i\left[\hat{H}_0 + \hat{H}_d(t), \rho\right]+\sum_k \kappa_k \mathcal{D}[\hat{d}_k] \rho
\end{equation}
\begin{equation}
d_{2}\rho= S_{k^*} \rho(t)+\rho(t) S_{k^*}^{\dagger}-\operatorname{tr}\left(S_{k^*} \rho(t)+\rho(t) S_{k^*}^{\dagger}\right) \rho(t)
\end{equation}
As mentioned previously, the $k^*$ indexed monitored operator $S_{k^*}$ represents the resonator transitions for the different qubit states (blue lines in Fig.~\ref{fig:bath_modes}). By construction, the stochastic master equation will produce the same average dynamics as the deterministic master equation, while providing realistic measurement trajectories.
Since the starting state of our readout simulation is always a pure state (eigenstate of qubit 0,1), we use QuTiP's stochastic Schrodinger equation solver $ssesolve$ \cite{qutip}.
Iteratively choosing QuTiP stochastic solver parameters \cite{qutip} for convergence, we find that about $ntraj \approx $12,800 trajectories, $nsubsteps \approx $200 substeps for a given time step of $1/(5\omega_d)$ worked well. The resulting numerical heterodyne measurement trajectories are demodulated by multiplying a phase $\exp{(i\omega_d t)}$ and integrated to generate Fig.~4 of main text, from which SNR and readout fidelity can be obtained.

\section{\label{sec:approx_in_ME_sim}Approximations in the master equation}
Master equations are derived with \cite{GardinerZoller} many approximations such as the Markovian or Born approximations. Our quartonic readout designs use very large $\kappa/2\pi = 300 $ MHz, so it is worth checking if these approximations are still valid.

In the literature, Purcell filters routinely have quality factors of about 30 and many hundreds of MHz of decay rate $\kappa_f$ (e.g. \cite{Sunada2023} $\kappa_f/2\pi =310$ MHz and \cite{PurcellFilter2014} $\kappa_f/2\pi=224$ MHz), and the standard Lindblad master equation was successful in reproducing experimental results \cite{Sete2015}. Therefore, the standard approximations used to derive the master equation such as the Markovian and Born approximations should still be valid in for our design with similar $\kappa/2\pi = 300 $ MHz and $Q = \omega_a / \kappa \approx 50$. 

However, our Purcell filter has much larger $\kappa_f = 4 \kappa = 2\pi \times 1.2 $ GHz decay rate, corresponding to a low quality factor of $Q \approx 13$. Although this is the same order of magnitude as experimentally demonstrations \cite{Sunada2023, PurcellFilter2014}, it is an open question whether the coupling of this Purcell filter to the bath is strong enough to violate the master equation's underlying assumptions. Therefore, we have opted to not include the Purcell filter as part of the quantum system in the master equation simulations, but have instead treated it more classically as a filtering function on the resonator's decay to the bath ($\kappa \propto [1+(2(\omega_{ji} - \omega_r)/\kappa_f)^2]^{-1}$).
Note also that we chose a small ratio of $\kappa_f/\kappa = 4$ but this has been experimentally demonstrated \cite{heinsoo2018}.

\section{\label{sec:resistor_loss}Loss from normal metals}

\begin{figure*}
    \centering
    \includegraphics[width=6in]{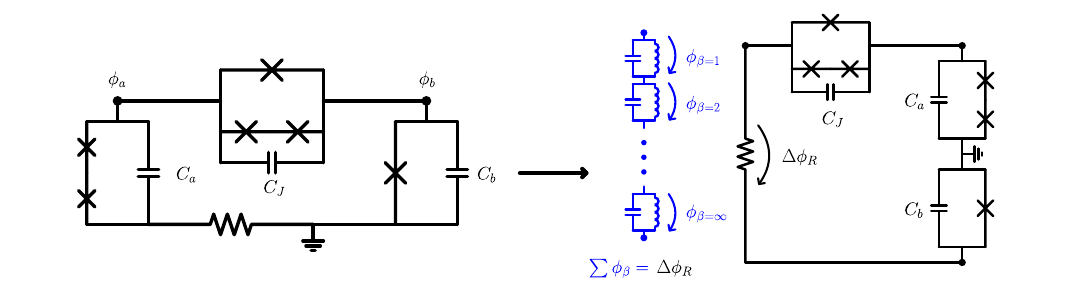}
    \caption{Rearrangement of the readout coupling circuit and illustration of the Caldeira-Legget model of the resistor to follow dissipative calculations in \cite{caldeiralegget}.}
    \label{fig:resist}
\end{figure*}

We include normal metal segments in our circuit as illustrated in Fig.~\ref{fig:resist} in order to dissipate DC flux offsets. The dissipation due to these resistors can be made small through the nonlinear nature of the quarton. Here, we follow the methodology introduced in Ref. \cite{caldeiralegget} to calculate the decoherence rate due to resistive dissipation from normal metal segments with a small resistance $R$. 

We can model the resistor Hamiltonian (let $\Phi_0/2\pi=1$) with the Caldeira-Legget model as
\begin{equation}
    H_R=\sum_\beta \left(\frac12C_\beta \dot\phi_\beta^2+\frac1{2L_\beta}\phi_\beta^2\right)
\end{equation}
with a total phase difference $\Delta\phi_R=\sum_\beta \phi_\beta$.
Here, we assume that the resistance is perturbative and much smaller than the reactance of the rest of the system, so $\phi_R=\sum_\beta\phi_\beta\ll1$ and that the quarton junction capacitances $C_J$ are much smaller than the resonator and qubit capacitances. Our new Lagrangian (allowing for general $\{n_S, n_{Ja}\}$ as in Eq.~(\ref{eq:H_diff_n})) will be 
\begin{equation}
    \begin{aligned}
        \mathcal L&=\frac12C_a\dot\phi_a^2+\frac12C_b\dot\phi_b^2+\frac12C_J(\dot\phi_a-\dot\phi_b-\Delta\dot\phi_R)^2\\
        &+n_{Ja}E_{Ja}\cos(\phi_a/n_{Ja})+E_{Jb}\cos(\phi_b)\\
        &+\alpha E_{J}\cos(\phi_a-\phi_b-\Delta\phi_R)-n_SE_{J}\cos\left(\frac{\phi_a-\phi_b-\Delta\phi_R}{n_S}\right)\\
        & +\mathcal L_R
    \end{aligned}
\end{equation}
which we can expand to first order in $\Delta\phi_R,\Delta\dot\phi_R$ so that 
\begin{equation}
    \begin{aligned}
    \mathcal L&=\mathcal L_0+\mathcal L_R-C_J(\dot\phi_a-\dot\phi_b)\Delta\dot\phi_R \\ 
    &+\alpha E_{J}\sin(\phi_a-\phi_b)\Delta\phi_R
    -n_SE_{J}\sin\left(\frac{\phi_a-\phi_b}{n_S}\right)\frac{\Delta\phi_R}{n_S}
    \end{aligned}
\end{equation}
where $\mathcal L_0$ is the original Lagrangian. We also assume that $\Delta\phi_R$ is small enough that we don't need to include it in our Legendre transformation. This gives us the modified Hamiltonian
\begin{equation}
    \begin{aligned}
    H & =H_0+H_R+\alpha E_{J}\sin(\hat\phi_a-\hat\phi_b)\Delta\phi_R \\
    & -E_{J}\sin\left(\frac{\hat\phi_a-\hat\phi_b}{n_S}\right)\Delta\phi_R\\
    &-C_J\biggl[\left(\frac{\hat q_a}{C_\Sigma/(C_b+C_J)}+\frac{\hat q_b}{C_\Sigma/C_J}\right)\\
    & -\left(\frac{\hat q_a}{C_\Sigma/C_J}+\frac{\hat q_b}{C_\Sigma/(C_a+C_J)}\right)\biggr]\Delta\dot\phi_R
    \end{aligned}
\end{equation}
where $C_\Sigma=C_aC_b+C_JC_a+C_JC_b$ and $H_0$ is the Hamiltonian without the resistor (see Eq.~(\ref{eq:H_diff_n})) . 

Treating the resistive terms with nonzero $\phi_R,\dot\phi_R$ as perturbations, we can apply Fermi's golden rule to obtain the transition probability from state $|i\rangle\to|f\rangle$ with $\Delta\phi_R,\Delta \dot\phi_R$ as uncorrelated noise terms. Using the transition formula $\Gamma=\frac{|\langle i|\hat A|f\rangle|^2}{\hbar^2}S_{FF}(\omega)$ for operator $\hat A$ and noise source $F$ from Ref. \cite{fermispectral}, we obtain our transition rate of:
\begin{widetext}
\begin{equation}
    \begin{gathered}
\Gamma=\Big|\langle i|2eC_J\left(\left(\frac{\hat n_a}{C_\Sigma/(C_b+C_J)}+\frac{\hat n_b}{C_\Sigma/C_J}\right)-\left(\frac{\hat n_a}{C_\Sigma/C_J}+\frac{\hat n_b}{C_\Sigma/(C_a+C_J)}\right)\right)|f\rangle\Big|^2S_{VV}(\omega_{if})/\hbar^2\\
+\Big|\langle i|\alpha E_{J}\sin(\hat\phi_a-\hat\phi_b)-E_{J}\sin\left(\frac{\hat\phi_a-\hat\phi_b}{n_S}\right)|f\rangle\Big|^2S_{\phi\phi}(\omega_{if}/\hbar^2.
\end{gathered}
\end{equation}

Following the derivations in \cite{caldeiralegget}, we can find the spectral densities for $\dot\phi_R=V_R,\phi_R$ as
\begin{equation}
    \begin{aligned}
    & S_{V_RV_R}(\omega)=\hbar\omega R\frac2{1-e^{-\hbar\omega/k_BT}}, \\
    & S_{\phi\phi}(\omega)=\frac{\hbar R}\omega\frac2{1-e^{-\beta\hbar\omega/k_BT}}\left(\frac{2\pi}{\Phi_0}\right)^2
    \end{aligned}
\end{equation}
and we take the limit of $T\to0$ in our cold superconducting environments so $S_{VV}(\omega)=2R\hbar\omega,S_{\phi\phi}(\omega)=\frac{2R\hbar}\omega\left(\frac{2\pi}{\Phi_0}\right)^2$. This gives our qubit dissipation rates as
\begin{equation}\label{eq:resistloss_supp}
    \begin{gathered}
    \Gamma=\frac{8e^2R\omega_b}{\hbar}\Big|\langle 01|\left(\frac{C_J}{C_a+C_JC_b/(C_b+C_J)}\hat n_a+\frac{C_J}{C_b+C_JC_a/(C_a+C_J)}\hat n_b-\frac{C_J^2}{C_\Sigma}(\hat n_a-\hat n_b)\right)|00\rangle\Big|^2\\
+\frac{8e^2R}\hbar\frac{(E_{J}/\hbar)^2}{\omega_b}\Big|\langle 01|\left(\frac{\alpha E_{J}}{E_{J}}\sin\left(\hat\phi_a-\hat\phi_b\right)-\sin\left(\frac{\hat\phi_a-\hat\phi_b}{n_S}\right)\right)|00\rangle\Big|^2\\
=\Gamma_C+\Gamma_Q.
    \end{gathered}
\end{equation}
\end{widetext}
The second term $\Gamma_Q$ in Eq.~(\ref{eq:resistloss_supp}) represents the resistive dissipation from energy flowing through the quarton. This term demonstrates why tilting the quarton helps reduce the resistive loss, where the tilt is defined as $t=2\alpha $. Heuristically, if we assume a small phase difference $\phi_a-\phi_b$ with some variance $\Delta\phi^2=\langle(\hat\phi_a-\hat\phi_b)^2\rangle$, expanding the operator in the second term yields
\begin{equation*}
\begin{aligned}
    &\frac{\alpha E_{J}}{E_{J}}\sin\left(\hat\phi_a-\hat\phi_b\right)-\sin\left(\frac{\hat\phi_a-\hat\phi_b}{n_S}\right) \\
    & \approx\frac{t}{n_S}\left(1-\frac{\Delta\phi^2}{6}\right)(\hat\phi_a-\hat\phi_b)-\frac1{n_S}\left(1-\frac{\Delta\phi^2}{6n_S^2}\right) (\hat\phi_a-\hat\phi_b)\\
    & =\left(t-\left(1+\frac{\Delta\phi^2}{6}\frac{n_S^2-1}{n_S^2}\right)\right)\frac{\hat\phi_a-\hat \phi_b}{n_S}
    \end{aligned}
\end{equation*}
which shows that the resistive is minimized for some $t>1$.

In our design, we don't need to tilt the quarton to minimize the resistive loss as long as the resistive dissipation isn't the limiting factor in our $T_1$ values. Instead, we can aim to minimize the linear coupling between the resonator and qubit to reduce Purcell decay. Using the parameters in Table \ref{tab:params}, Eq.~(\ref{eq:resistloss_supp}) evaluates to $1/\Gamma\approx 0.113\text{ s}$. Here, $1/\Gamma_C=0.871\text{ s},1/\Gamma_Q=0.130\text{ s}$.

\section{\label{sec:flux_noise}Flux noise analysis}

\begin{figure}
    % \centering
    \includegraphics[width=0.5\textwidth]{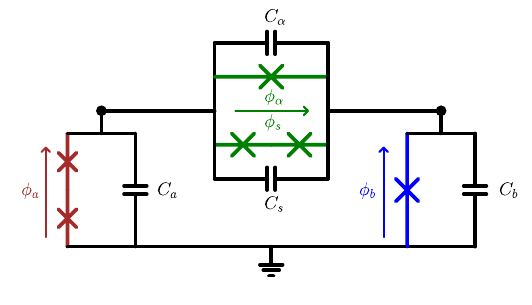}
    \caption{Branch variables used in flux noise analysis}
    \label{fig: flux_supp}
\end{figure}

To estimate decoherence induced by flux noise, we follow the approach of \cite{time_dep_flux} to write an ``irrotational" Hamiltonian for our system, eliminating the term proportional to the time derivative of the flux noise. We assume that the flux noise through the two loops is independent, so we can analyze noise in the two loops separately. This means we only ever have one time dependent flux, simplifying the analysis. 

We define branch fluxes and capacitances as in Fig.~\ref{fig: flux_supp}. For time-dependent flux $\tilde\phi = 2\pi\frac{\Phi}{\Phi_0}$ in the ground loop, this yields

\begin{equation}
    \begin{gathered}
    \hat{H}_{irr} = T - 2E_{Ja}\cos\big(\frac{1}{2}(\phi_a - \frac{C_a^{-1}}{\sum_iC_i^{-1}}\tilde\phi)\big) \\
    - E_{Jb}\cos\big(\phi_b + \frac{C_b^{-1}}{\sum_iC_i^{-1}}\tilde\phi\big) \\
    + E_{J\alpha}\cos\big(\phi_b-\phi_a-\frac{C_Q^{-1}}{\sum_iC_i^{-1}}\tilde\phi\big) \\
    -2E_{Js}\cos\big(\frac{1}{2}(\phi_b-\phi_a-\frac{C_Q^{-1}}{\sum_iC_i^{-1}}\tilde\phi\big),
    \end{gathered}
\end{equation}
where $T$ is the usual capacitive energy term and $C_Q = C_s + C_\alpha$. Similarly, if the time-dependent noise $\tilde\phi$ threads the quarton loop, we obtain
\begin{equation}
    \begin{gathered}
    \hat{H}_{irr} = T - 2E_{Ja}\cos\big(\frac{1}{2}(\phi_a + A\tilde\phi)\big) - E_{Jb}\cos\big(\phi_b + B\tilde\phi\big) \\
    + E_{J\alpha}\cos\big(\phi_b-\phi_a + (B-A)\tilde\phi\big) \\
    -2E_{Js}\cos\big(\frac{1}{2}(\phi_b-\phi_a +(B-A+1)\tilde\phi\big),
    \end{gathered}
\end{equation}
with 
\begin{equation}
    A = \frac{C_b C_s}{2C_a C_b + (C_a + C_b)(C_\alpha + C_s)}
\end{equation}
\begin{equation}
    B = \frac{C_a C_s}{2C_a C_b + (C_a + C_b)(C_\alpha + C_s)}
\end{equation}
We can then compute the $T_1$ decoherence contribution of each loop as

\begin{equation}
    \Gamma_{\Phi}=\left|\bra{0}\frac{\partial H}{\partial \Phi}\ket{1}\right|^2 S_{\Phi}\left(\omega_q\right)
\end{equation}
with $S_{\Phi}(\omega)=A_{\Phi}^2\left(\frac{2 \pi \times 1 \mathrm{~Hz}}{\omega}\right)^{\gamma_{\Phi}}$. We use $A_{\Phi}^2 = \left(1 \: \mu \Phi_0\right)^2 / \mathrm{Hz}$ and $\gamma_\Phi = 1$ for the quarton loop \cite{EngineersReview}. Since we anticipate the ground loop being larger, we increase the flux noise amplitude to $A_{\Phi}^2 = \left(5 \: \mu \Phi_0\right)^2 / \mathrm{Hz}$, corresponding to a 25x increase in loop perimeter \cite{flux_noise_geometry}. Note that although the normal metal resistor precludes DC flux bias in the ground loop, the LR time constant for the loop is too slow to prevent noise at the qubit frequency. 

When estimating the pure dephasing caused by flux noise, we choose to simulate an echo measurement rather than a Ramsey measurement. A Ramsey measurement would be very sensitive to the length of numerical time series used (as a longer time series includes more low frequency noise). We simulate the echo measurement by first generating a long ($\approx$ 3 hour) time series with power spectral density given by $S_{\Phi}(\omega)$, and then dividing it into many short time series. This avoids artificially filtering out low frequency noise that could affect the echo measurement. For each flux noise time series, we compute the qubit frequency at each time and integrate it (with an added sign flip in the middle) to compute the dephasing of a typical echo sequence. We then average the resulting echo sequence over all the time series.

\bibliography{scibib}% Produces the bibliography via BibTeX.

\end{document}